\documentclass[a4paper,fleqn,usenatbib]{mnras}

\usepackage{newtxtext,newtxmath}
\usepackage[T1]{fontenc}
\usepackage{ae,aecompl}

\usepackage{graphicx}	% Including figure files
\usepackage{amsmath}	% Advanced maths commands
\usepackage{amssymb}	% Extra maths symbols
\usepackage[usenames]{color}

\usepackage{upgreek}
\usepackage{aas_macros} % required to get journal abbreviations
\usepackage{url}
\usepackage{booktabs}

\title[g~mode frequency groups in KIC~5608334]{An astrophysical interpretation of the remarkable g-mode frequency groups of the rapidly rotating $\gamma$\,Dor star, KIC~5608334}

\author[H. Saio et al.]{
Hideyuki Saio$^{1}$\thanks{E-mail: saio@astr.tohoku.ac.jp (HS)},
Timothy R. Bedding$^{2,3}$,
Donald W. Kurtz$^{4}$,
Simon J. Murphy$^{2,3}$,
\newauthor{
Victoria L. Antoci$^{3}$,
 Hiromoto Shibahashi$^{5}$,
 Gang Li$^{2,3}$,
and Masao Takata$^{5}$}
\\
$^{1}$Astronomical Institute, Graduate School of Science, Tohoku University, Sendai, Miyagi 980-8578, Japan\\
$^{2}$Sydney Institute for Astronomy, School of Physics, The University of Sydney, NSW 2006, Australia\\
$^{3}$Stellar Astrophysics Centre, Department of Physics and Astronomy, Aarhus University, Ny Munkegade 120, DK-8000 Aarhus C, Denmark\\
$^{4}$Jeremiah Horrocks Institute, University of Central Lancashire, Preston PR1 2HE, UK\\
$^{5}$Department of Astronomy, School of Science, The University of Tokyo, Bunkyo-ku, Tokyo 113-0033, Japan \\
}

\date{Accepted XXX. Received YYY; in original form ZZZ}

\pubyear{2018}

\begin{document}
\label{firstpage}
\pagerange{\pageref{firstpage}--\pageref{lastpage}}
\maketitle

\begin{abstract}
The Fourier spectrum of the $\gamma$-Dor variable KIC~5608334 shows remarkable frequency groups at $\sim$3, $\sim$6, $\sim$9, and 11--12\,d$^{-1}$. We explain the four frequency groups as prograde sectoral g~modes in a rapidly rotating star. Frequencies of intermediate-to-high radial order prograde sectoral g~modes in a rapidly rotating star are proportional to $|m|$ (i.e., $\nu \propto |m|$) in the co-rotating frame as well as  in the inertial frame. This property is consistent with the frequency groups of KIC~5608334 as well as the period vs. period-spacing relation present within each frequency group, if we assume a rotation frequency of  $2.2$\,d$^{-1}$, and that each frequency group consists of prograde sectoral g~modes of $|m| = 1, 2, 3,$ and 4, respectively. In addition, these modes naturally satisfy near-resonance conditions $\nu_i\approx\nu_j+\nu_k$ with $m_i=m_j+m_k$. 
We even find exact resonance frequency conditions (within the precise measurement uncertainties) in many cases, which correspond to combination frequencies.
\end{abstract}


\section{Introduction}

Alan Cousins (1903 -- 2001) remarkably published in this journal for 77 years. His first paper, on observations of the light curve of the Cepheid $\ell$\,Carinae \citep{cousins1924}, was published in 1924, and his last, on photometric extinction \citep{cousins2001}, was published on the day he died, 2001 May 11 \citep{kilkenny2001}. 

Cousins first became interested in the light variation of $\gamma$\,Doradus at least as early as the 1960s when \citet{cousins1963} reported variability in $\gamma$\,Doradus with a range in photographic magnitude of 0.04 mag; they gave the variability type as ``I?'', meaning indeterminate. They noted that some of the observations of the stars in their paper dated to before 1952. So the original mystery of the light variability of $\gamma$\,Doradus began in the middle of the last century. Stimulated by Cousins' work, further observations were made in the late 1960s by \citet{stobie1971}, who noted that $\gamma$\,Doradus has a period in the range $0.33 - 1.00$\,d, and that it might be a $\beta$\,Lyrae or W~Ursa~Majoris star with shallow eclipses. Interestingly, from the modern $\upmu$mag perspective of the {\it Kepler} mission data, the title of Stobie's paper was ``Microvariability of bright A and F stars'', where hundredths of a magnitude variation, and mmag precision were state-of-the-art. 

By the 1980s Cousins had found that $\gamma$\,Doradus was at least doubly periodic (\citealt{cousins1989},  \citealt{cousins1992}, \citealt{cousins1994}), but he was still noting that the ``cause of the variation is not known''. He had a fascination with this star, and talked to his many colleagues about it, including Kurtz and Balona. Kurtz performed a frequency analysis of Cousins' data for $\gamma$\,Doradus in collaboration with him, but made no progress; Balona did the same and was successful. The big breakthrough came when \citet{balona1994} showed that two principal frequencies in $\gamma$\,Doradus are stable and phase-locked, and they found evidence of a third frequency. They ruled out starspots as the source of the variability, and concluded that ``this star is the best example of what appears to be a new class of pulsating F-type variables.'' 

Thus was born the class of $\gamma$\,Dor stars, which we now know are multi-periodic g-mode pulsators. Many studies followed over the next two decades. But those studies were plagued by what \citet{balona1994} referred to as an ``aperiodic component'' to the light variations. The second breakthrough came with data of unprecedented precision and duration with the {\it Kepler} space mission. With those data we now know that the $\gamma$\,Dor stars have many g\:modes of consecutive radial order whose frequencies are so closely spaced that data spanning at least a few months are needed to resolve them. With the pulsation frequencies of $\gamma$\,Dor stars typically being in the $0-4$\,d$^{-1}$ range, ground-based observations are inadequate to resolve the daily alias confusion for these stars. It is simply not possible to come even close to obtaining continuous data for months, and impossible to obtain continuous data for years from the ground, as the {\it Kepler} mission did from space. Our understanding of the $\gamma$\,Dor stars is an unintended consequence (benefit!) of a space mission built for an entirely different purpose -- the search for Earth-like exoplanets \citep{bor10}. 

The $\gamma$\,Dor stars are of fundamental importance to our understanding of stellar structure and evolution because the g~modes probe the core conditions of these stars. Since the 1960s g~modes have been sought in the Sun for this purpose, but without success that is universally accepted 
 (\citealt{app10}, although see \citealt{fos17}). 
For the $\gamma$\,Dor stars there is no doubt: we are probing the core conditions from just above the convective energy generation zone, right out to the stellar surface for ``hybrid'' stars that also show $\delta$~Sct p-mode pulsations, and those hybrids are abundant in the {\it Kepler} data set. 

Of particular interest is our new ability to study the internal rotation of stars in detail during their main-sequence, hydrogen-burning phase. For some of the many observational studies now addressing this, see \citealt{vanr16}, \citealt{mur16}, \citealt{schmid2015}, \citealt{vanreeth2015}, \citealt{sai15} and \citealt{kur14}. For fascinating theoretical discussions of the diagnostic abilities of the g\,modes for $\gamma$\,Dor stars, see \citet{oua17} and \citet{bou13}.

We now understand that the observed `aperiodicity' in the light curves of $\gamma$\,Dor stars is actually closely spaced series of g-mode frequencies. Nevertheless, problems remain in understanding the light curves of $\gamma$\,Dor stars, and the related $\delta$~Sct stars, as well as other A stars that do not show any pulsational variability \citep{mur15}. 

\citet{kur15} provided a unifying explanation for a variety of light curve shapes among $\gamma$\,Dor, Slowly Pulsating B (SPB) and pulsating Be stars in terms of combination frequencies based on only a few pulsation modes. They particularly addressed the stars described by \citet{mcnamara2012} as having frequency groups (fg), and found that combination frequencies of a few base frequencies in the principal group could explain all of the peaks in the other frequency groups. Previous attempts had been made to extract frequencies from the groups and treat them all as pulsation mode frequencies, but \citet{kur15} suggested no need for that. Yet harmonics and combination frequencies arise from highly non-linear pulsation, and \citet{kur15} gave no explanation of why some $\gamma$\,Dor and SPB stars should show such strong non-linearity, while other stars do not. 

\begin{figure}
\includegraphics[width=\columnwidth]{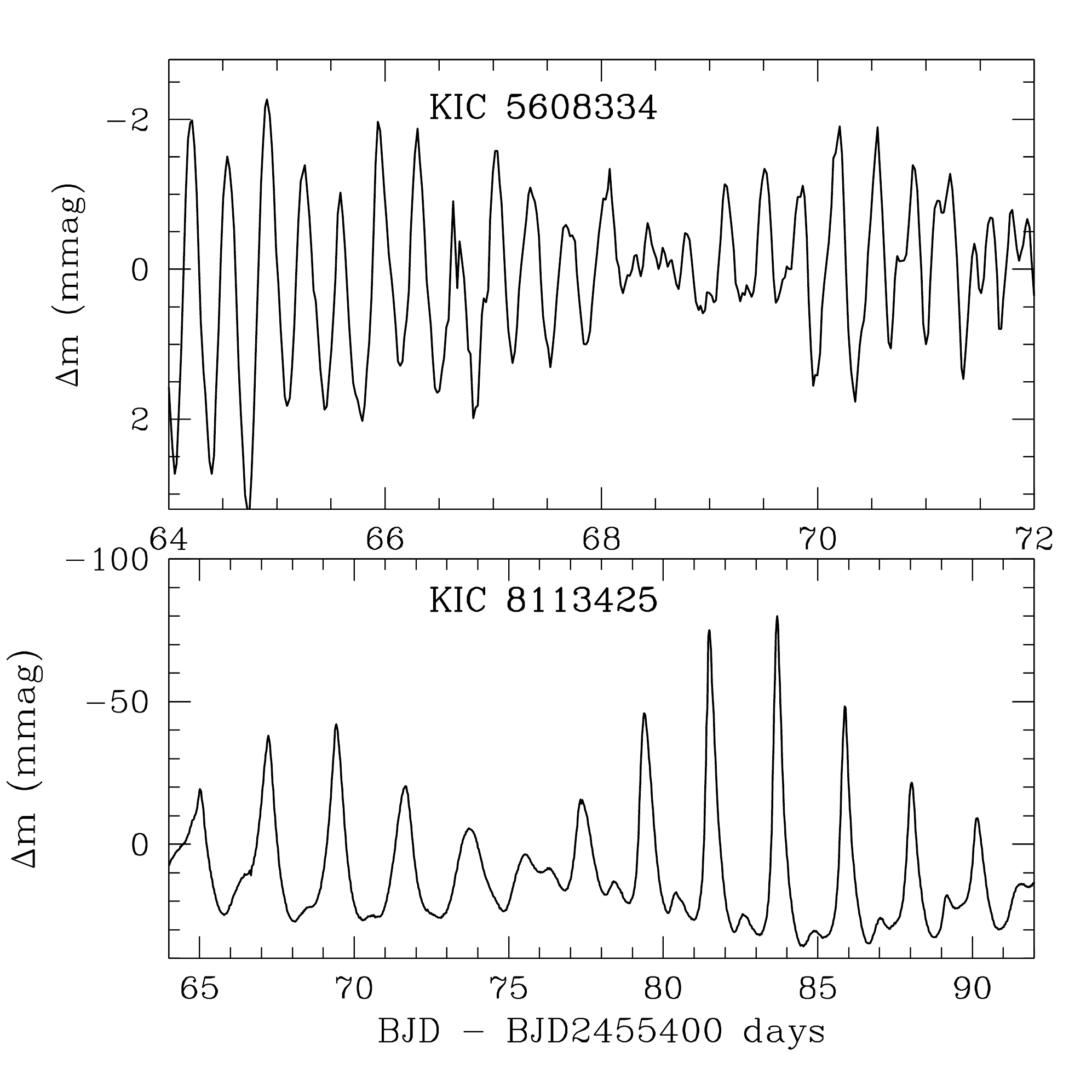} %{k5608334_k8113425_LLC.pdf}
\caption{A section of the long-cadence light curves of KIC~5608334 (top) and KIC~8113425 (bottom). The oscillations of KIC\,5608334 are linear and small in amplitude, whereas KIC\,8113425 has highly non-linear oscillations of much larger amplitude.}
\label{fig:LC}
\end{figure}

In this paper we discuss how rapid rotation can produce frequency groups similar to those discussed in \citet{kur15} even for relatively small amplitude pulsators (i.e. with weak non-linearity), taking the $\gamma$ Dor star KIC~5608334 as an example. 
Fig.~\ref{fig:LC} compares portions of the Kepler light curves of  KIC~5608334 and KIC~8113425. The latter star is one of the $\gamma$ Dor stars discussed by \citet{kur15}. Obviously, the amplitude of KIC~8113425 is much larger and the light curve has a strongly non-linear nature with asymmetric positive and negative excursions, while the light curve of KIC~5608334 is symmetrical. Still, the amplitude spectrum of KIC~5608334 shows strong frequency groupings (Fig.~\ref{fig:ft} below) similar to those of  KIC~8113425 \citep{kur15}.

We suggest that the frequency groups of g modes appear in rapidly rotating stars, in which the rotational shift of prograde sectoral modes  of consecutive degree ($-m =  1, 2, 3, 4, \ldots$) generates mode frequencies that are very close to the harmonics and combination frequencies of the base mode frequencies. Resonance then 
causes
the pulsation mode frequencies in the frequency groups to exactly match the combination frequencies. It is noteworthy that detailed pulsation models provide a good description of the pulsation mode frequencies in the frequency groups of the {\it Kepler} $\gamma$\,Dor star KIC\,5608334, as we show in this paper. 

This hypothesis gives an astrophysical reason why some stars show frequency groups and others do not, and it is testable by measurement of $v \sin i$ in a large ensemble of $\gamma$\,Dor stars, both with and without frequency groups. Because of the relative faintness of the {\it Kepler} stars, observations to get accurate $v \sin i$ are challenging, but they can be made. The primary goal of this paper is to describe models for KIC\,5608334 for prograde sectoral pulsations with $-m = 1, 2, 3, 4, \ldots$, and to show how they match the observations. 

In a non-rotating star, the angular dependence of a nonradial pulsation mode is designated by integers $\ell$ and $m$ of a spherical harmonic $Y_\ell^m$. The distribution of radial displacement (and variations of scaler quantities) has no latitudinal nodal line if $\ell=|m|$, these are called sectoral modes, while in the other cases, $(\ell-|m|)$  latitudinal  nodal lines appear and those are called tesseral modes \citep[see e.g.,][]{unno,ack10}. In a rotating star, in particular if the rotation frequency is larger than the pulsation frequency in the co-rotating frame, a single $Y_\ell^m$ cannot be used to describe a pulsation mode because a mixing among different $\ell$ occurs. Still,  to describe the property of the amplitude distribution on the stellar surface,  we use the adjectives `sectoral' and `tesseral'  for non-axisymmetric modes without and with latitudinal nodal lines, respectively. Sometimes, we use in this paper `the first tesseral mode' to indicate a mode with one latitudinal nodal line.

\section{Model}

Equilibrium main-sequence models to obtain theoretical pulsation frequencies were calculated using Modules for Experiments in Stellar evolution \citep[MESA;][]{pax13} in the same way as our previous works on $\gamma$\,Dor stars \citep{kur14,sai15,mur16}. We have adopted a standard chemical composition of $(X,Z) = (0.72,0.014)$ with the OPAL opacity tables \citep{igl96}, and the mixing-length is set to be $1.7H_{\rm p}$, with $H_{\rm p}$ being the pressure scale height. The effects of the Coriolis force on the pulsation frequencies are included non-perturbatively using the method of \citet{lee95}, where the effect of centrifugal deformation is included approximately to the second order of angular rotation frequency. The latter assumption is justified because g~modes propagate in the deep interior so that the effects of deformation on the g-mode frequencies are small \citep{bal12}. In the method of \citet{lee95}, to calculate pulsation frequencies in a rotating star, eigenfunctions are expanded into terms proportional to spherical harmonics. We truncated the expansion at the 6th ($\pm 1$ depending the convergence of eigenfunctions) term. All the theoretical frequencies used in this paper were obtained under the adiabatic approximation.

\section{KIC~5608334 -- a rapidly rotating $\gamma$ Dor star}

KIC~5608334 is a $\gamma$\,Dor variable of spectral type F2\,V \citep{nie15}. At \mbox{$V = 9.9$\,mag} it is relatively bright compared to most \textit{Kepler} $\gamma$\,Dor stars, which allowed \citet{nie15} to observe it at high spectral resolution. The spectroscopic parameters they obtained are listed in Table\,\ref{tab:param}. GAIA DR1 \citep{gaia16}  gives a parallax of  $3.035\pm 0.385$\,mas. The parallax, combined with a bolometric correction \citep{flo96}, yields the luminosity of KIC~5608334 listed in Table\,\ref{tab:param}. 

\begin{figure}
\includegraphics[width=\columnwidth]{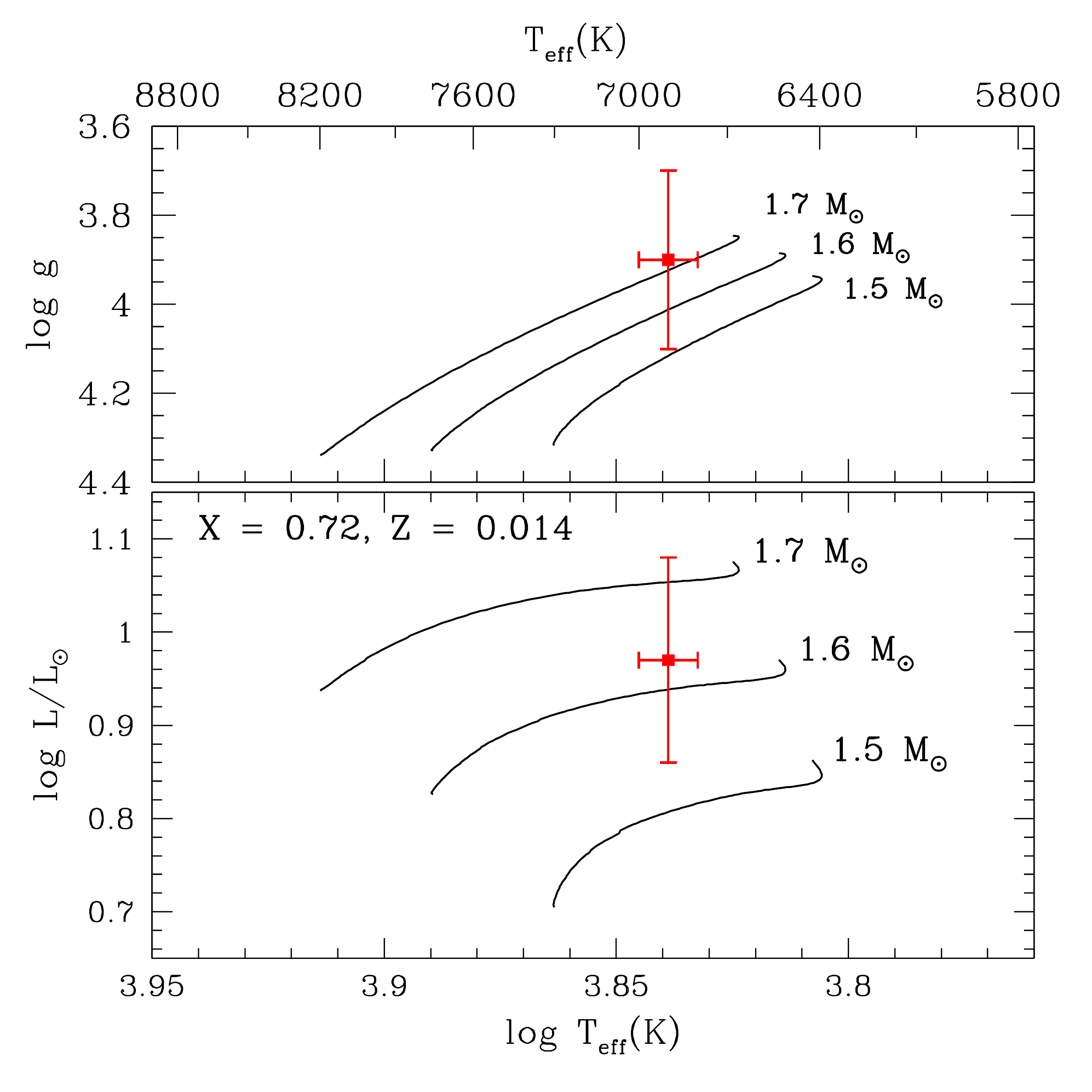}  %{hrd_teg_m1p5tom1p7.pdf}
 \caption{Some evolutionary tracks and estimated positions of KIC~5608334 (see Table~\ref{tab:param}) with error bars in the HR diagram (bottom panel) and the $\log T_{\rm eff}-\log g$ diagram (top panel). Effects of rotation are not included in the evolutionary models. The luminosity of KIC~5608334 was obtained from the GAIA (DR1) parallax. 
 }
\label{fig:hrd}
\end{figure}

The positions of KIC~5608334 in the HR diagram and the $\log T_{\rm eff}$--$\log g$ diagram are shown in Fig.\,\ref{fig:hrd} with some evolutionary tracks for a normal composition $(X=0.72,Z=0.014)$, which is consistent with the spectroscopy. The estimated luminosity is roughly consistent with the spectroscopic surface gravity, $\log g$, indicating a mass range of $1.5 - 1.7$\,M$_\odot$. To examine the pulsation properties of  KIC~5608334, we adopted models in this mass range having effective temperatures consistent with the spectroscopic range as listed in   Table\,\ref{tab:param}.

\begin{table}
	\centering
	\caption{Parameters for KIC~5608334; the spectroscopic parameters are from \citet{nie15}, the luminosity is calculated from the GAIA DR1 \citep{gaia16} parallax.}
	\label{tab:param}
	\begin{tabular}{lcr@{\,$\pm$\,}l}
		\toprule
		Parameter & Unit & \multicolumn{2}{c}{Value} \\
		\midrule
		$\log T_{\rm eff}$  & (K) & 3.839 & 0.006 \\
		$T_{\rm eff}$  & K & 6900 &100 \\
		$\log g$  & (cgs) & 3.9 & 0.2 \\
		$v \sin i$ & km\,s$^{-1}$ & 110 & 13\\
		 {\rm [Fe/H]} & & $-0.05$ & 0.12\\
		\midrule
		$\log L/{\rm L}_\odot$ & &  0.97 & 0.11 \\
		\bottomrule
	\end{tabular}
\end{table}

\begin{figure*}
\includegraphics[width=\textwidth]{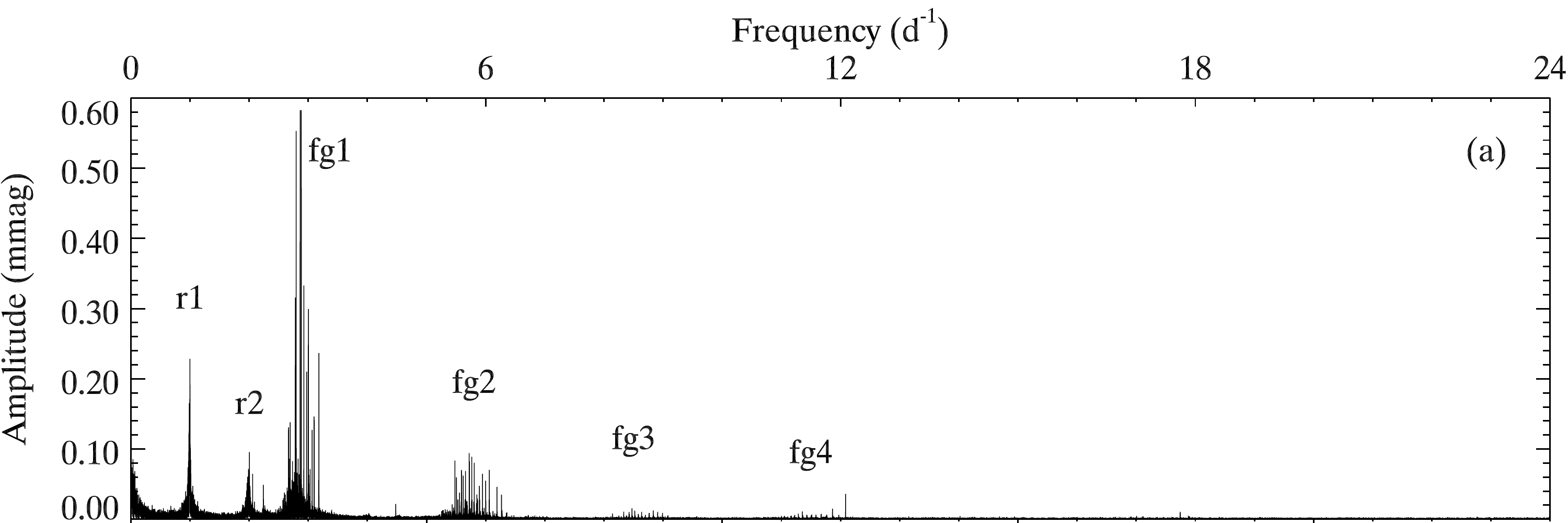}  %{kic5608334_FT_0,24.pdf}
\includegraphics[width=\textwidth]{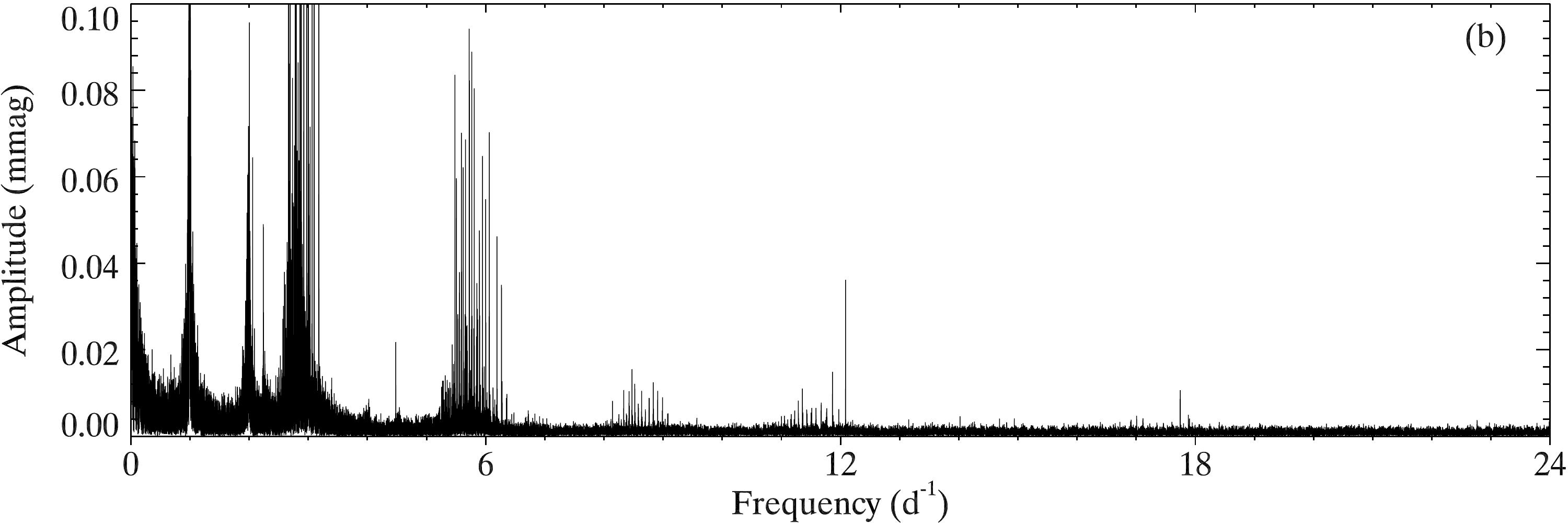}  %{kic5608334_FT_0,24_zoom.pdf}
\caption{ The amplitude spectrum obtained from 1470\,d of Kepler long cadence light curves of KIC~5608334 nearly out to the Nyquist frequency for long cadence data. 
The two panels have different vertical scales.
Four frequency groups are identified. Abbreviating frequency group as fg, we refer to them as fg1, fg2,  fg3, and  fg4, as indicated in the top panel. Lower frequency groups designated as r1 and r2 are considered to be r~modes, as discussed in \citet{sai18}. 
In this paper we associate the frequency groups with  prograde sectoral g~modes of azimuthal order $-m= 1,  2,  3$, and $4$, respectively.}
\label{fig:ft}
\end{figure*}

Fig.~\ref{fig:ft} shows the amplitude spectrum of the full 1470-d Kepler light curve of KIC~5608334. We identify four frequency groups (labelled fg) in the ranges fg1: 2.7--3.2\,d$^{-1}$,  fg2: $5.3 - 6.4$\,d$^{-1}$,   fg3: $8.1 - 9.2$\,d$^{-1}$ and  fg4: $11.0 - 12.2$\,d$^{-1}$. It is remarkable that frequencies of fg2, fg3 and fg4 are in the ranges, respectively, of twice,  three times, and four times that of fg1. We identify these frequency groups fg1 $\ldots$ fg4 as prograde sectoral g~modes of $-m= 1, 2, 3$, and $4$, respectively. (In this paper we adopt the convention that a negative $m$ corresponds to a prograde mode.)  Lower frequency groups r1 at $\sim\!1$\,d$^{-1}$ and r2 at $\sim\!2$\,d$^{-1}$ are considered to be r~modes, as discussed in \citet{sai18}.

Fig.~\ref{fig:ft} shows the presence of a peak at 2.2397~d$^{-1}$ (and the harmonic at 4.479~d$^{-1}$). We consider this peak the rotation frequency at a surface spot. The frequency is slightly higher than the rotation frequency $2.20$~d$^{-1}$ determined in \S\ref{sec:pspace} by comparing the g-mode period spacings of KIC~5608334 with models (where uniform rotation is assumed). The closeness  of the two frequencies implies that the star rotates almost uniformly, although the slight difference, if significant, indicates the presence of a slight latitudinal and/or radial differential rotation.

\subsection{Pulsation frequencies}

We have downloaded the  long cadence SAP (simple aperture photometry) data of KIC~5608334 from the KASOC (Kepler Asteroseismic Science Operations Center) web site (http://kasoc.phys.au.dk/index.php) as ascii files. In order to account for the different zero points from quarter to quarter we simply 
divided the fluxes in each quarter by their median and then converted to parts per million [ppm]. 
Oscillation frequencies of KIC~5608334 were measured from the full 1470-d Kepler light curve by using two different methods. As a first approach we used the software {\sc PERIOD04} \citep{period04}. For a more detailed frequency extraction, however, we employed automated software based on the classical iterative prewhitening process, where the highest peak in the Lomb-Scargle periodogram was identified and then subtracted from the light curve. The statistical significance of each peak was assessed  by using the false alarm probability \citep{sca82}, which gives good results in the case of the {\it Kepler} data. In addition, the amplitude of each extracted peak was compared to the value in the original un-prewhitened data, allowing a maximum deviation of 25\%. This step, which was also used by \citet{vanr15}, allowed us to make sure that the peak was not introduced while subtracting other signals. 
This software, which is based on the Timeseries Tools code \cite{timeseries_tools},  will be presented and discussed in more detail in an upcoming paper (Antoci et al., in prep.).

Employing the procedure described above, i.e., keeping the peaks with an amplitude ratio between the extracted and the original value in the range $0.75 - 1.00$, we found 66 significant peaks; however, only 36 are 
above 2\,d$^{-1}$ corresponding to the frequency groups fg1 -- fg4. The lower-frequency peaks in the groupings r1 and r2 (Fig.\,\ref{fig:ft}) are too closely spaced to be resolved, even with 4.0 years of {\it Kepler} data, so we disregard these values. 
To avoid introducing additional signals while prewhitening peaks, we filtered the data (simple high- and low-pass filtering) such that we can extract frequencies for each of the fg groupings individually. Applying this more elaborate procedure, we identified a total of 192 peaks satisfying the criteria described above. Those frequencies are listed in Table\,\ref{tab:freq} in Appendix.

\begin{figure*}
\begin{center}
 \includegraphics[width=\columnwidth]{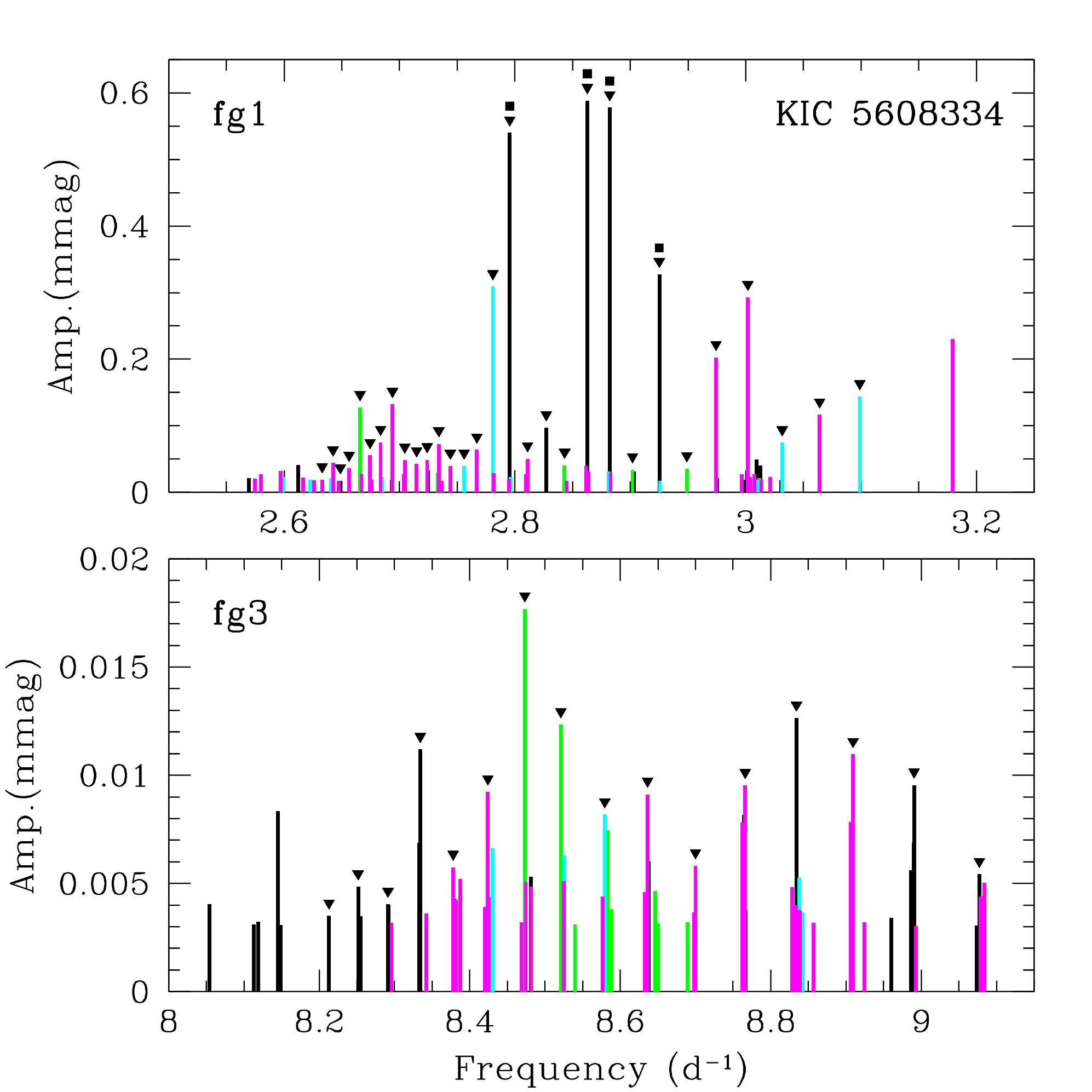}  %{freq_amp_comb_fg13.pdf}
\includegraphics[width=\columnwidth]{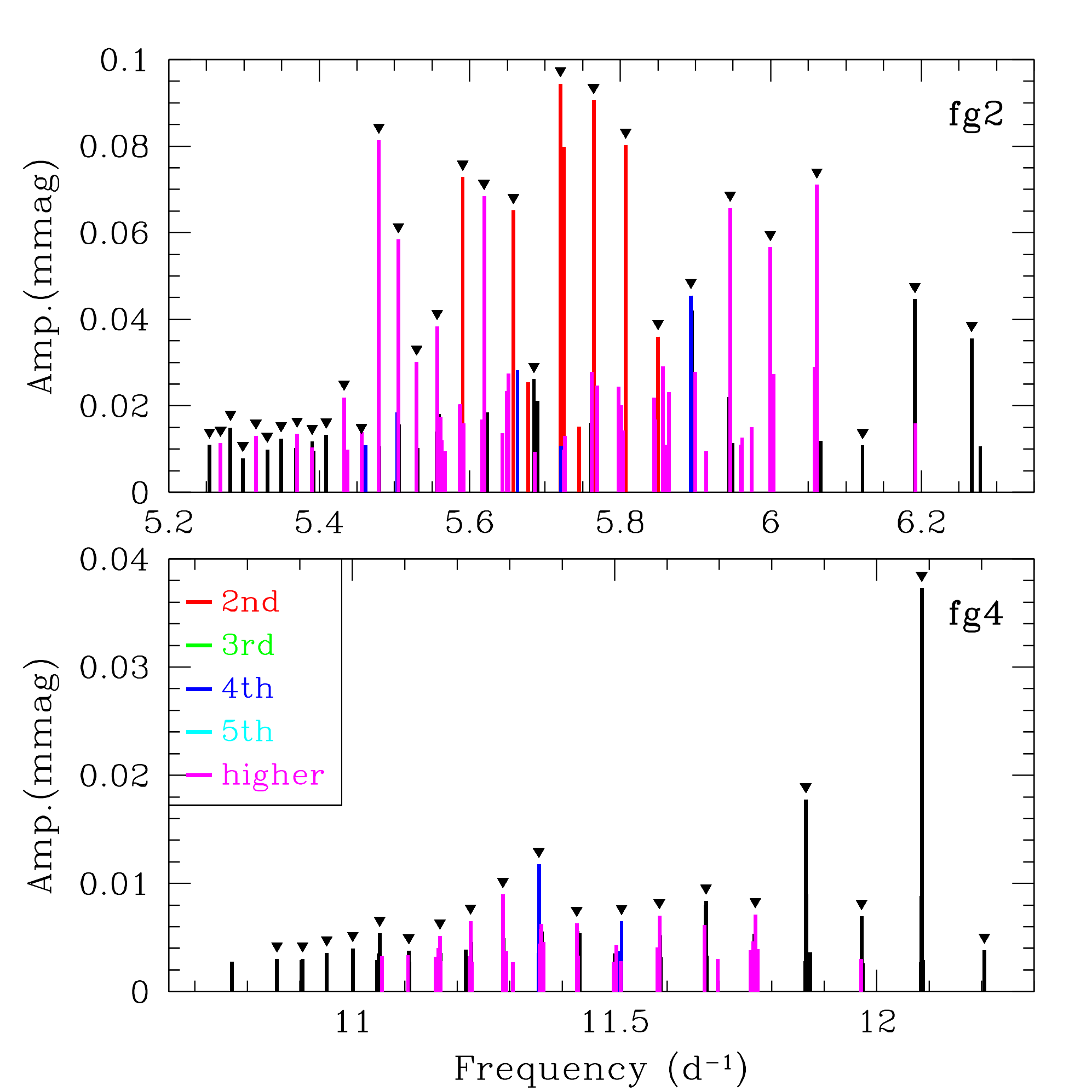}   %{freq_amp_comb_fg24.pdf}
\end{center}	
\caption{Amplitude spectra for each frequency group of KIC~5608334. We identify all combination frequencies,  using up to three of four parent modes (indicated by black full squares at tops in the upper left panel); i.e. $a\nu_i \pm b\nu_j \pm c\nu_k$ with positive integers $a$, $b$, and $c$. Combination frequencies of different orders are shown by different colours as explained in the legend in the lower right panel. The order refers to the sum of all coefficients; i.e., $a$+$b$+$c$. 
The inverted triangles indicate  frequencies used to calculate period spacings shown in Fig.\,\ref{fig:inert}. }
\label{fig:freq_amp}	
\end{figure*}

We searched for combination frequencies in the form $a\nu_i \pm b\nu_j \pm c\nu_k$, using up to three of the four frequencies of the highest amplitudes (indicated by filled black squares in Fig.\,\ref{fig:freq_amp}), where $a, b, c$ are integers satisfying the conditions, $0\le a,b,c \leq 4$ and $a+b+c \le 9$.
 A peak was identified as a combination frequency if the absolute value of the difference between the predicted combination frequency and the measured peak was lower than the resolution, i.e. $|\nu_{\rm combination} - \nu_{\rm obs}| < 1/\Delta T$, where $\Delta T = 1470$~d.\footnote{Although frequencies at large amplitude peaks may be measured more accurately, we adopt $1/\Delta T = 6.8\times10^{-4}$\,d$^{-1}$ as a conservative uncertainty of frequencies for the low-amplitude pulsator KIC~5608334.}  We found 69 combination frequencies, which are shown in Fig.~\ref{fig:freq_amp} with different colours depending on the order (i.e., $a+b+c$).
 We discuss, in the latter part of this paper, why eigenfrequencies of a rapidly rotating star are observed to be close to the combination frequencies.

\subsection{Period spacings of g~modes}\label{sec:pspace}

\begin{figure}
\includegraphics[width=\columnwidth]{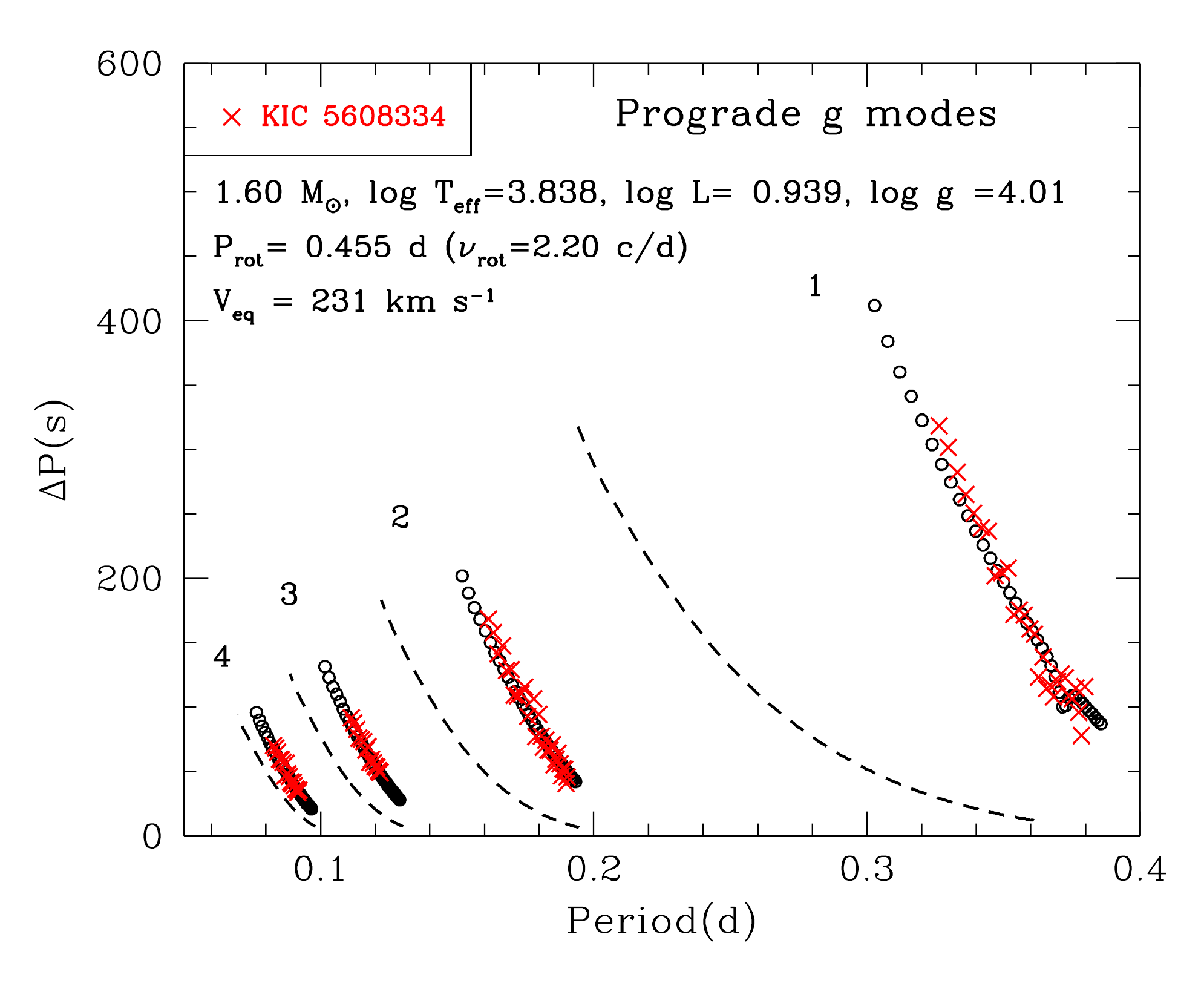}  %{gmodes_dp_p_inert_m1p6_2p20_k5608334VA.pdf}
\caption{Period spacings of KIC~5608334 (red crosses) are compared with theoretical predictions from a model of 1.6\,M$_\odot$ with a rotation frequency of $2.20$\,d$^{-1}$. The observed first to fourth frequency groups are fitted with prograde sectoral g~modes of $-m=1$, $2$, $3$ and $4$ (open circles), respectively. For each $m$, g~modes with radial orders from 21 to 60 are plotted. The dashed lines indicate the predicted relation for the first tesseral ($\ell-|m|=1$ in the non-rotating case) prograde g-mode sequence for each $m$. }
\label{fig:inert}
\end{figure}

Amplitude spectra of KIC~5608334  for each frequency group shown in Fig.~\ref{fig:freq_amp} indicate that the majority of frequencies are more-or-less regularly spaced.  Using the frequencies indicated by inverted triangles in Fig.~\ref{fig:freq_amp}, we have calculated period spacings ($\Delta P$), which are plotted (crosses) with model predictions for g modes (circles and dashed lines) in Fig.\,\ref{fig:inert} as a function of period. Within each frequency group, $\Delta P$ decreases with period, which is a common property of prograde g~modes of a rotating star as discussed in, e.g., \citet{bou13}, \citet{vanr16} and \citet{oua17}. The gradient is steeper for faster rotation so that we can determine the rotation frequency by fitting the gradient with models. 
We compared the gradients of period spacings of KIC~5608334 with theoretical ones for rotation rates of $2.24$\,d$^{-1}$ and $2.20$\,d$^{-1}$ (uniform rotation is assumed). Although a peak at $2.24$\,d$^{-1}$ appears in Fig.\,\ref{fig:ft}, we found that the rotation frequency of $2.20$\,d$^{-1}$ agrees with the period spacings of KIC~5608334 slightly better. Therefore, we have adopted $2.20$\,d$^{-1}$ for the rotation frequency of KIC~5608334. Fig.~\ref{fig:inert} compares 
theoretical $\Delta P$ values of prograde sectoral (open circles) and first tesseral (dashed lines) g~modes of  $-m= 1,  2, 3$ and $4$ of a 1.6-M$_\odot$ model rotating at a frequency of $2.20$\,d$^{-1}$. Prograde sectoral g~modes, rather than tesseral modes, reproduce well the properties of the $\Delta P$--period relations of KIC~5608334. 

Since the rotation frequency affects not only the gradient of the $\Delta P$--period relation, but also the prediction for the period (i.e. frequency) range of each frequency group,  the agreement of both quantities with a single rotation frequency strongly  supports our identification of the frequency groups of KIC~5608334 as prograde sectoral g~modes with different azimuthal orders $m$. We note that similar good agreements are obtained for models of 1.5-M$_\odot$ and 1.7-M$_\odot$ with similar $T_{\rm eff}$ as long as the rotation frequency  $2.20$\,d$^{-1}$ is assumed. While we recognise that many frequencies in higher frequency groups are combinations of the frequencies in fg1 (Fig.~\ref{fig:freq_amp}), the good agreement of our models of prograde sectoral g~modes with the observed frequency ranges gives an astrophysical basis for the existence of the g-mode frequency groups in a rapidly rotating star. 

\begin{figure*}
	\includegraphics[width=\columnwidth]{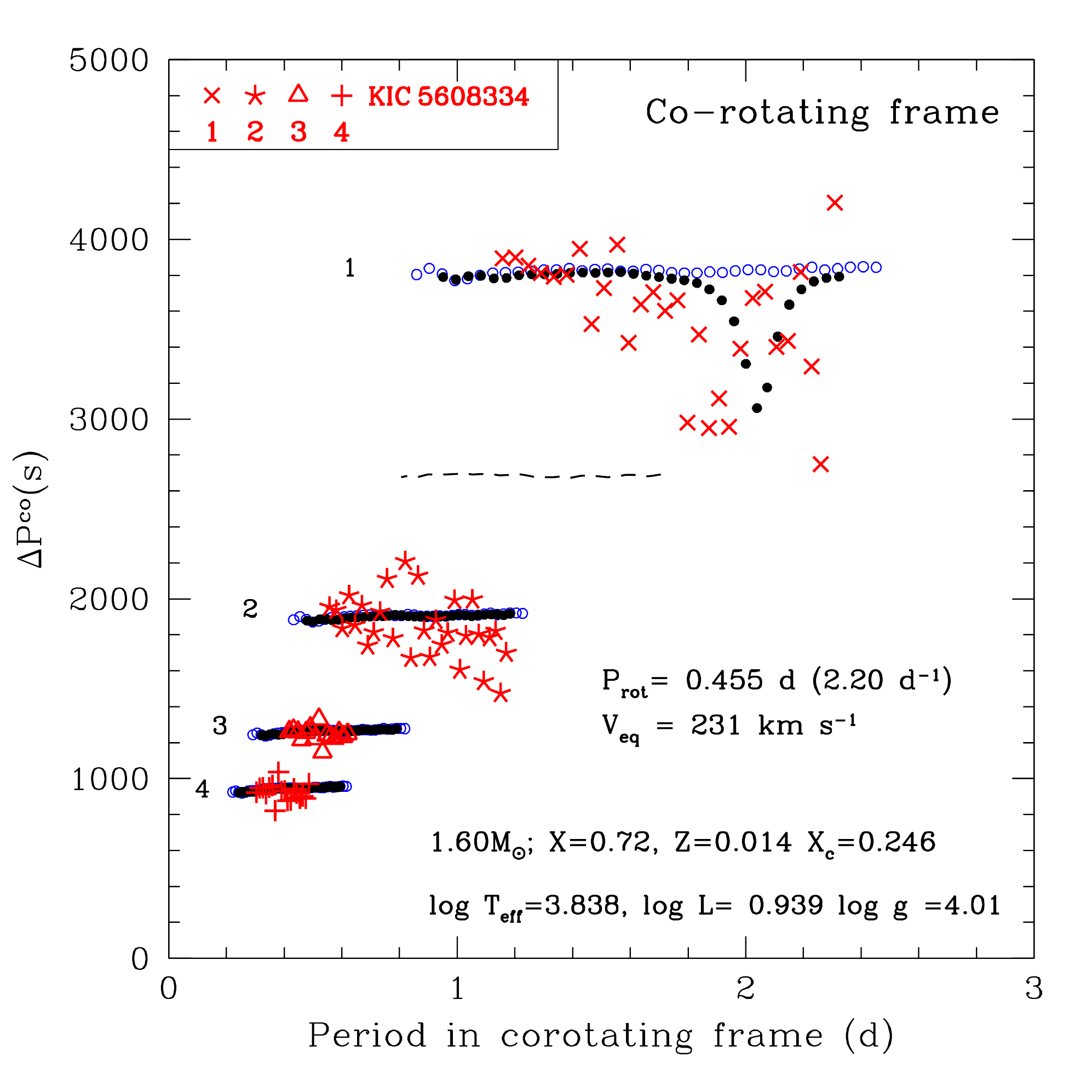}  %{pdpcorot_m1p60_13_2p20_VA.pdf}
	\includegraphics[width=\columnwidth]{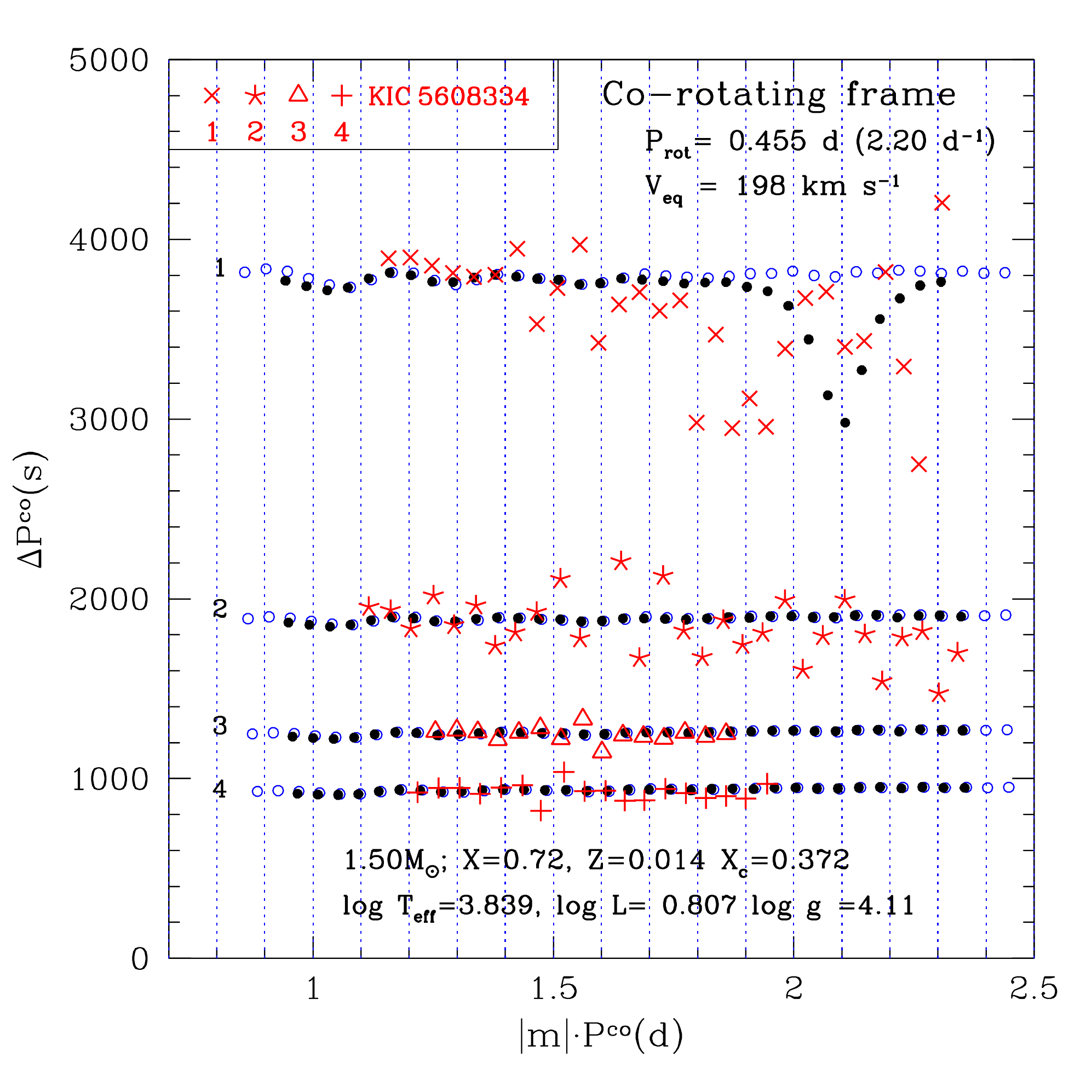}  %{pdpcorot_mp_m1p50_10_2p20_VA.pdf}
\caption{Period spacings of KIC~5608334 converted to the co-rotating frame assuming a rotation frequency of $2.2\,$d$^{-1}$, compared with theoretical values from 1.6-M$_\odot$ (left panel) and 1.5-M$_\odot$ (right panel) models.  (Different masses are used to show that model frequencies are insensitive to adopted masses.) Blue open circles are from calculations with the Traditional Approximation of Rotation (TAR), while black filled circles are results of full calculations without using the TAR. The ordinates of both panels are the same; they show period spacings of g modes in the co-rotating frame in seconds. The abscissa of the left panel is period in the co-rotating frame, $P^{\rm co}$, in days, while that in the right panel adopts $|m|$ times $P^{\rm co}$ to visualize resonance conditions, where the azimuthal orders of observed frequencies belonging to fg1, fg2, fg3 and fg4 are assumed to be $-m=1, 2, 3$, and $4$, respectively. Vertical dotted lines in the right panel are drawn at every 0.1~d (arbitrarily) for visibility of approximate alignments of frequencies.  For each azimuthal order $m$, g~modes with radial orders, $n_g$,  from $22$ to $54$ (i.e. left-to-right) are plotted ($20 \le n_g \le 56$ for the results from the TAR; blue circles). The dashed line in the left panel shows $\ell=1$ periods and period spacings in non-rotating case.
Observational points belonging to different frequency groups of KIC~5608334 are shown by different symbols; i.e., crosses: fg1; asterisks: fg2; triangles: fg3; and plusses:  fg4. }
\label{fig:corot}
\end{figure*}

Using the rotation frequency $\nu_{\rm rot} = 2.20$\,d$^{-1}$ and identifying the azimuthal order $m$ for each group of g-mode  frequencies of KIC~5608334, we can convert the detected frequencies to those in the co-rotating frame by subtracting $|m|\nu_{\rm rot}$. We can then compare period spacings $\Delta P^{\rm co}$ as a function of period in the co-rotating frame with our models. Fig.\,\ref{fig:corot} shows such comparisons with 1.6-M$_\odot$  (left panel) and  1.5-M$_\odot$  (right panel) models; the former model is the same as that in Fig.\,\ref{fig:inert}. The abscissa in the left column is period in the co-rotating frame, $P^{\rm co}$, and is $|m|P^{\rm co}$ in the right panel. We have adopted models of different masses between the left and the right panel to show that the agreement with observed properties is insensitive to stellar mass, as long as the same rotation frequency $2.2$~d$^{-1}$ is used. This is also consistent with the findings of  \citet{oua17}.

The theoretical period spacings in the co-rotating frame are nearly constant as a function of $P^{\rm co}$, with some wiggles that are caused by the hydrogen abundance profile just above the convective core \citep{mig08}. Nearly constant values of $\Delta P^{\rm co}$ indicate that the Coriolis force affects the g~modes strongly \citep[][\S\ref{sec:prop}]{bal12,bou13}.

The observational data roughly agree with the model predictions with relatively large scatter. The enhancement of the scatter is inevitable because  the quantity subtracted, $|m|\nu_{\rm rot}$, from each frequency in the inertial frame consists of a large fraction, which enhances the fractional uncertainties. The fact that the observational $\Delta P^{\rm co}$ roughly distribute horizontally supports our choice of rotation frequency, $2.2$~d$^{-1}$ for KIC~5608334. Periods and the period range for a larger $|m|$ are smaller (left panel of Fig.\,\ref{fig:corot}).  This tendency is compensated in the right panel by using an abscissa of $|m|P^{\rm co}$, in which prograde sectoral g~modes with the same radial order but different $|m|$ align vertically; we discuss the reason in  the next section. 

It is remarkable that the radial orders of g~modes corresponding to the observed periods are confined to a range between $\sim22$ and  $\sim54$, irrespective of the values of $|m|$ (i.e., irrespective of frequency groups). This property is consistent with resonance couplings among modes with different $m$ (discussed in Sec.~\ref{sec:resonance} below), and also consistent with the result of the nonadiabatic analysis for non-rotating models of $\gamma$ Dor stars by \citet{dup05}  that,  among g modes of different $\ell$,  modes with similar ranges of radial orders are excited.  Probably, both effects contribute to the property.

Blue open circles in Fig.\,\ref{fig:corot} show results obtained using the Traditional Approximation of Rotation (TAR), in which the horizontal component of the angular velocity of rotation is neglected. The approximation generally produces accurate results for low-frequency nonradial pulsations, in which horizontal motions dominate. This fact is seen in this figure, agreeing in general  with the results of full computations (filled black circles). However, there is an appreciable difference in period spacings of $m=-1$ sectoral g~modes, where there is a dip in the full calculations but not in the calculations with the TAR. That dip seems to be caused by a very weak coupling between a sectoral mode and a tesseral mode. Such a coupling never occurs under the TAR. Interestingly, the observed period spacings seem to suggest the presence of such a dip in the period spacings for the first group.

The period spacing of $\ell=1$ g~modes in the non-rotating model (horizontal dashed line in Fig.\,\ref{fig:corot}) is smaller than that in the co-rotating frame of $m=-1$ sectoral modes in the rotating model. This is because the effective latitudinal degree of prograde sectoral modes decreases with rotation. For the same reason, prograde sectoral g~modes of higher $\lvert m \rvert$ have smaller period spacings. Such properties will be discussed in the next section. 

By comparing the period spacings of KIC~5608334 with models, we determined its rotation frequency to be $2.2$~d$^{-1}$ irrespective to an assumed mass, while the corresponding equatorial rotation velocity $V_{\rm eq}$ depends on the radius of a model. 
At $T_{\rm eff}=6900$\,K, the $1.5-$M$_\odot$ and the $1.7-$M$_\odot$ models have radii of  $1.78$\,R$_\odot$ and $2.37$\,R$_\odot$, respectively, which correspond to $V_{\rm eq}=198$~km\,s$^{-1}$ and $264$~km\,s$^{-1}$. 
 From $v\sin i$ given in Table~1, we estimate a 1$\sigma$ range of inclination of the rotation axis from $22^\circ$ to $38^\circ$. 

\section{Properties of Low-frequency g-mode oscillations of a rotating star} 
\label{sec:prop}

In the presence of rotation, the latitudinal degree $\ell$ cannot be specified for a pulsation mode, because a pulsational perturbation proportional to a spherical harmonic $Y_\ell^m(\theta,\phi)$ is not independent of a perturbation proportional to $Y_{\ell'}^m$ with $\ell'\not=\ell$ due to the effects of the Coriolis force and centrifugal deformation \citep[e.g.,][]{unno,ack10}. This complicates significantly the calculation of pulsation modes in a rotating star, requiring two-dimensional calculations \citep[e.g.,][]{ree09} or expansion of eigenfunctions with multiple spherical harmonics \citep{lee95}.

The Traditional Approximation of Rotation (TAR) is useful, in particular, for understanding properties of low-frequency pulsations in a rotating star, in which pulsation frequencies in the co-rotating frame are comparable to, or lower than, the rotation frequency. In this approximation, the horizontal component of angular velocity of rotation ($\Omega\sin\theta$, with $\theta$ being co-latitude) is neglected. As Fig.\,\ref{fig:corot} indicates, the TAR is generally a good approximation for the low-frequency pulsations in a rotating star.  Here, we discuss qualitative properties of such low-frequency pulsations using this approximation.

In the TAR, a set of equations for non-radial pulsations under the Cowling approximation (which neglects the Eulerian perturbation of the gravitational potential) is preserved, except that $\ell(\ell+1)$ is replaced with $\lambda$, the eigenvalue of Laplace's tidal equation, which depends on the ratio of the rotation  frequency, $\nu_{\rm rot}$, to the pulsation frequency in the corotating frame, $\nu^{\rm co}$. We can use the asymptotic formulae of high-order g~modes in non-rotating stars for g~modes in rotating stars if $\ell(\ell+1)$ is replaced with $\lambda$. Thus, the frequency of a high-radial-order g~mode in a rotating star can be represented as
\begin{equation}
    \nu^{\rm co} \approx {\sqrt{\lambda} \over 2\pi^2n_{\rm g}}\int {N\over r} dr  
    \equiv {\sqrt{\lambda}\over n_g}\nu_0,
\label{eq:freqco}
\end{equation}
\citep{lee87b,bou13} where $N$ is the Brunt-V\"ais\"al\"a frequency, $n_g$ is the radial order of the g mode, and $\nu_0$ is a frequency defined as above. (This equation is also applicable to r~modes,  as discussed by \citealt{sai18}.) Although the apparent form of the equation is very similar to the  non-rotating case, variation of $\lambda$ as a function of $2\nu_{\rm rot}/\nu^{\rm co}$ (= spin parameter) generates properties substantially different from those of non-rotating stars.

In a slowly rotating star  $\lambda$ is given as
\begin{equation}
\lambda \approx \ell(\ell+1) + m{2 \nu_{\rm rot}\over\nu^{\rm co}},  \quad {\rm if} \quad 2\nu_{\rm rot}/\nu^{\rm co} \ll 1
\end{equation}
\citep{ber78}, while if $2\nu_{\rm rot}/\nu^{\rm co} > 1$,
the value of  $\lambda$ for 
g~modes becomes drastically different from $\ell(\ell+1)$:
\begin{equation}
\left. 
\begin{array}{ll}
\lambda \approx m^2; \quad \mbox{prograde sectoral g~modes} \cr
\lambda\propto\left({2\nu_{\rm rot}\over\nu^{\rm co}}\right)^2 \gg m^2 ~; ~ \mbox{other g~modes} \cr
\end{array}
\right\}
~{\rm if}  \quad {2\nu_{\rm rot}\over\nu^{\rm co}}  > 1,
\label{eq:lambda}
\end{equation} 
 \citep[see, e.g.,][]{lee97,sai17}; i.e., $\lambda$ of prograde sectoral g~modes decreases from $\ell(\ell+1)$ to $m^2$ with increasing spin parameter, while $\lambda$ of retrograde or tesseral g~modes increases rapidly and becomes much larger than $m^2$. 

Substituting the above expressions for $\lambda$ into equation (\ref{eq:freqco}), we obtain
\begin{equation}
\left. 
\begin{array}{ll}\displaystyle
\nu^{\rm co} \approx {|m|\nu_0\over n_g} ~; \quad \mbox{prograde sectoral g  modes} \cr \\  \displaystyle
\nu^{\rm co} > \sqrt{2\nu_{\rm rot}\nu_0\over n_g} ~; \quad \mbox{other g  modes} \cr
\end{array}
\right\}
~{\rm  if } ~ {2\nu_{\rm rot}\over\nu^{\rm co}} >1. 
\label{eq:nuco}
\end{equation} 
Inverting the relation for a prograde sectoral g~mode leads to a relation of $n_g/\nu_0 \approx |m|P^{\rm co}$, which explains the vertical alignment of modes with the same radial order but with different $|m|$ in the right panel of Fig.\,\ref{fig:corot}.
We note that for all frequency groups of KIC~5608334,  spin parameters ($=2\nu_{\rm rot}/\nu^{\rm co}$) are always larger than unity. They are  $12 - 4.5$ for fg1, $5.2- 2.3$ for fg2, $3.0 - 1.8$ for fg3, and $2.1 - 1.3$ for fg4.

From equation (\ref{eq:nuco}) we can express period spacing of prograde sectoral modes in the co-rotating frame as 
\begin{equation}
\Delta P^{\rm co} \approx {1\over |m|\nu_0}; ~\mbox{prograde sectoral g~modes};
\label{eq:Dp}
\end{equation}
i.e., $\Delta P^{\rm co}$ is approximately constant and the value is proportional to $1/|m|$. 
 This is the property of model predictions we see in Fig.\,\ref{fig:corot}, which is roughly supported by the observational data of KIC~5608334. If the modes in  KIC~5608334 were  tesseral, $\Delta P^{\rm co}$ would be much smaller and systematically change as $\propto 1/P^{\rm co}$, which is not consistent with the observations.

We note here that in the non-rotating case, equation~(\ref{eq:Dp}) corresponds to the equation $\Delta P\approx [\sqrt{\ell(\ell+1)}\nu_0]^{-1}$. Because $\ell=|m|$ for sectoral modes in the non-rotating case, non-rotating period spacings (the horizontal dashed line in the left panel of Fig.\,\ref{fig:corot}) are always smaller than those of prograde sectoral modes, $\Delta P^{\rm co}$, in the rotating case. 

\subsection{Properties in the inertial (observational) frame}

Adopting the convention that a negative $m$ corresponds to a prograde mode, pulsation frequency in the inertial frame is written as
\begin{equation}
\nu^{\rm int} = \nu^{\rm co} - m\nu_{\rm rot} = {\sqrt{\lambda}\over n_g}\nu_0 - m\nu_{\rm rot},
\label{eq:freq_int_co}
\end{equation}
where the last equality applies for g~modes.
Using the property of $\lambda$ in equation~(\ref{eq:lambda}) we obtain for prograde sectoral g~modes
\begin{equation}
\nu^{\rm int}_{n_g} \approx |m|\left({\nu_0\over n_g} + \nu_{\rm rot}\right) ;\quad \mbox{prograde sectoral g~modes},
\label{eq:freqsecint}
\end{equation}
if $2\nu_{\rm rot} > \nu^{\rm co}$.
Thus, the frequencies of prograde sectoral g~modes in the inertial frame are proportional to $|m|$. This property explains the frequency grouping of KIC\,5608334 seen in Fig.\,\ref{fig:ft}.   To see how well the relation is satisfied, we list, in Table\,\ref{tab:modelfreq}, samples of prograde sectoral g~modes of $m=-2$ and $-4$ to compare them with  $2\times$ and $4\times$ the corresponding $m=-1$ prograde sectoral g-mode frequencies obtained {\it without} using the TAR, in which the same 1.6-M$_\odot$ model as in Fig.\,\ref{fig:inert} was adopted. These numbers indicate that the proportionality relation given in equation (\ref{eq:freqsecint}) is satisfied well in the model. Thus, the frequency groupings of KIC\,5608334 can be explained by the property of low of frequency prograde sectoral g~modes with different azimuthal order $m$ influenced by rapid rotation. 

\begin{table}
	\centering
	\caption{Examples of theoretical frequencies (d$^{-1}$) in the inertial frame for sectoral g~modes of $m=-1$ (column 2) and twice (column 3) and four times  (column 5) in comparison with corresponding frequencies of $m=-2$ (column 4), and $-4$ (column 6), respectively. Although these frequencies are obtained by full calculations without the TAR, these numbers have the property represented by equation\,(\ref{eq:freqsecint}) based on the TAR.
	 }
	\label{tab:modelfreq}
	\begin{tabular}{cccccc} 
		\toprule
		(1)   &  (2)       &    (3)       &   (4)       &   (5)       &   (6)     \\
		$n_g$ &  $ m=-1$  & $2\times(2)$  &$m = -2$ & $4\times(2)$ & $m=-4$ \\
		\midrule
		60 &   2.58618 & 5.17236 & 5.15970 & 10.3447 & 10.3144 \\
		39 &   2.78835 & 5.57670 & 5.57354 & 11.1534 & 11.1309 \\ 
		38 &   2.80397 & 5.60794 & 5.60469 & 11.2159 & 11.1920 \\  
        37 &   2.82050 & 5.64100 & 5.63761 & 11.2820 & 11.2564 \\ 
        24 &   3.16228 & 6.32456 & 6.31302 & 12.6491 & 12.5568 \\
		\bottomrule
	\end{tabular}
\end{table}

Using equation (\ref{eq:freqsecint}), we can estimate observational period spacings of prograde sectoral g~modes as
\begin{equation}
\Delta P^{\rm int} ={1\over\nu^{\rm int}_{n_g+1}}-{1\over\nu^{\rm int}_{n_g}} \approx 
{1\over |m|}{\nu_0\over(\nu_0+n_g\nu_{\rm rot})^2} ~ ,
\end{equation}
where $n_g\gg1$ is assumed.
This indicates that the period spacing of prograde sectoral g~modes in the inertial frame  decreases with radial order (i.e., with increasing period) for a given $|m|$ (i.e., within a frequency group), while for a given radial order $n_g$  the period spacing decreases with $|m|$. This explains the properties seen in Fig.~\ref{fig:inert}. 

\subsection{Amplitude distribution on the surface}

\begin{figure}
\includegraphics[width=\columnwidth]{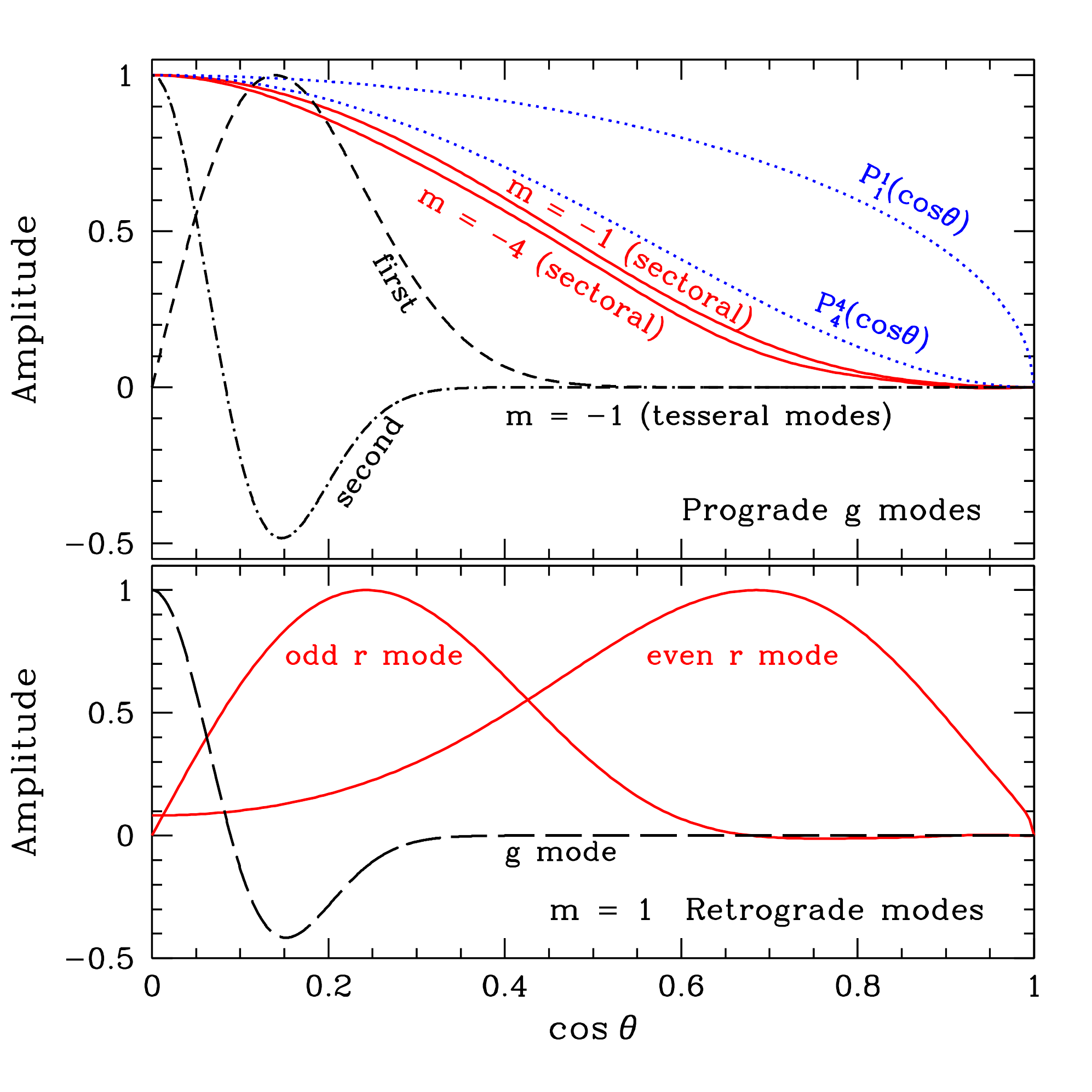}   %{amp_dist.pdf}
\caption{Radial part of the displacement (or temperature variation) amplitude across the stellar surface (hemisphere) of selected modes, with $\theta$ being co-latitude (i.e. $\cos\theta=0$ corresponds to the equator). The amplitude of each mode is normalised so that the maximum is unity. The 1.6-M$_\odot$ model is used with a rotation frequency of $2.2$\,d$^{-1}$. Upper and lower panels are for prograde and retrograde (in the co-rotating frame) modes, respectively. The sectoral prograde modes of $m=-1$ and $-4$ (solid lines in the top panel) have a radial order $n_g=35$.
 The odd and even r~modes in the bottom panel have inertial frame frequencies of  $1.0$\,d$^{-1}$ ($n_g=53$) and $2.0$\,d$^{-1}$ ($n_g=35$) (Fig.\,\ref{fig:ft}; see \citet{sai18} for the property of r modes). 
The  prograde tesseral g~modes in the top panel and a retrograde `sectoral' g~mode were obtained by using the TAR, because without this approximation strong interactions with other modes prevent us from obtaining a target mode. For these modes $2\nu_{\rm rot}/\nu^{\rm co} = 6.7$ is assumed; the prograde sectoral mode of $m=-1$ shown in the top panel has a similar value. Dashed and dash-dotted lines in the upper panel are the first and the second tesseral modes, respectively, which correspond to $\ell=2$ and $\ell=3$ at $\nu_{\rm rot}=0$, respectively.  Dotted lines in the top panel are the Legendre functions $P_1^1(\cos\theta)$ and $P_4^4(\cos\theta)$, the amplitude distributions for $\ell=|m|=1$ and $4$ modes in non-rotating stars. }
\label{fig:amp}
\end{figure}

Rotation generally concentrates the pulsation amplitude of a g~mode toward the equator (Fig.\,\ref{fig:amp}; see also Fig.\,\ref{fig:patterns} for 3D graphics). The effect is stronger for tesseral modes and retrograde modes. For retrograde g~modes, additional latitudinal nodal lines appear if $2\nu_{\rm rot}/\nu^{\rm co}  > 1$. Therefore, a retrograde sectoral g~mode of $m = \ell$ becomes a tesseral mode by the addition of latitudinal nodal lines (in both the north and south hemispheres)  if $2\nu_{\rm rot}/\nu^{\rm co}  > 1$; i.e., {\bf no sectoral {\it retrograde} g~modes are expected in a rapidly rotating star.}

Fig.\,\ref{fig:amp}  shows that among g~modes, the amplitudes of prograde sectoral modes are less affected by rotation, thus should have highest visibility. The latitudinal distribution of the $m=-4$ prograde sectoral modes is less affected by rotation and is comparable to that of the $m=-1$ prograde sectoral modes of KIC\,5608334, because  $\nu^{\rm co}$ of the $m=-4$ prograde sectoral modes are higher by a factor of four than that of $m=-1$ prograde sectoral modes. Although the latitudinal distribution is similar,  the visibility of $m=-4$ modes should be much less than that of $m=-1$ because of the azimuthal variation of the amplitude, $\sin(m\phi)$. According to \citet{das02} the visibility ratio between $\ell = 4$ and $\ell = 1$ is $\sim0.03$, while the amplitude ratio of the fourth group to the first group of KIC\,5608334 is roughly $0.02$, indicating that $m=-4$ prograde sectoral modes are excited to intrinsic amplitudes comparable to $m=-1$ prograde sectoral modes, and the difference in observed surface amplitudes is largely geometric in origin. (A similar argument holds for $-m=2,3$, though those seem to be smaller by factors of two or three.)

Fig.\,\ref{fig:A2} shows  the distribution of temperature variations (colour coded) and horizontal displacements (arrows) on the surface for the g-mode pulsation in the middle of each frequency group of KIC~5608334. Horizontal displacements are mainly azimuthal in the case of  a large spin parameter.  

\section{Two- or three-mode resonance couplings }
\label{sec:resonance}

A non-linear two- or three-mode coupling among $i,j,k$ modes (two-mode coupling if $j=k$) occurs if
\begin{equation}
\begin{array}{ll}
m_i = m_j + m_k  \cr
{\rm and} \cr 
\nu^{\rm co}_i = \nu^{\rm co}_j + \nu^{\rm co}_k + \delta\nu \quad {\rm with} \quad |\delta\nu|\ll\nu^{\rm co}_i.\cr
\end{array}
\label{eq:resonance}
\end{equation}
Here $m_i$ and $\nu^{\rm co}_i$ are the azimuthal order and the linear frequency in the co-rotating frame of mode $i$, respectively.
Representing the pulsation as $\Re[A_a\boldsymbol{\xi}_a\exp(2\pi {\rm i}\nu_a^{\rm co}t)]$ with $a  = i, j, k$ ,
we obtain an amplitude equation \citep[cf.][]{dzi82} % derived an amplitude equation
\begin{equation}
{{\rm d}A_i\over {\rm d}t} = \gamma_iA_i+{\rm i}\alpha_iA_jA_k{\rm e}^{-2\pi {\rm i}(\delta\nu) t} 
\label{eq:A_i}
\end{equation}
and two similar equations for $dA_j/dt$ and $dA_k/dt$.
Here, $\gamma_i$ is the linear growth rate of the linear pulsation mode $i$, 
and $\alpha_i$ 
represents the strength of the non-linear coupling
\citep[the detailed form of coupling is discussed by, e.g.,][]{dzi82}. 
If $A_j A_k$ in the second term of the right hand side of equation (\ref{eq:A_i}) is roughly constant, and if the typical value of the second term is much larger than the linear excitation/damping term represented by the first term, then $A_i$ is proportional to $\exp[-2\pi {\rm i} (\delta\nu)t]$. Then the oscillation with the combination frequency is realised.

\begin{figure}
\includegraphics[width=\columnwidth]{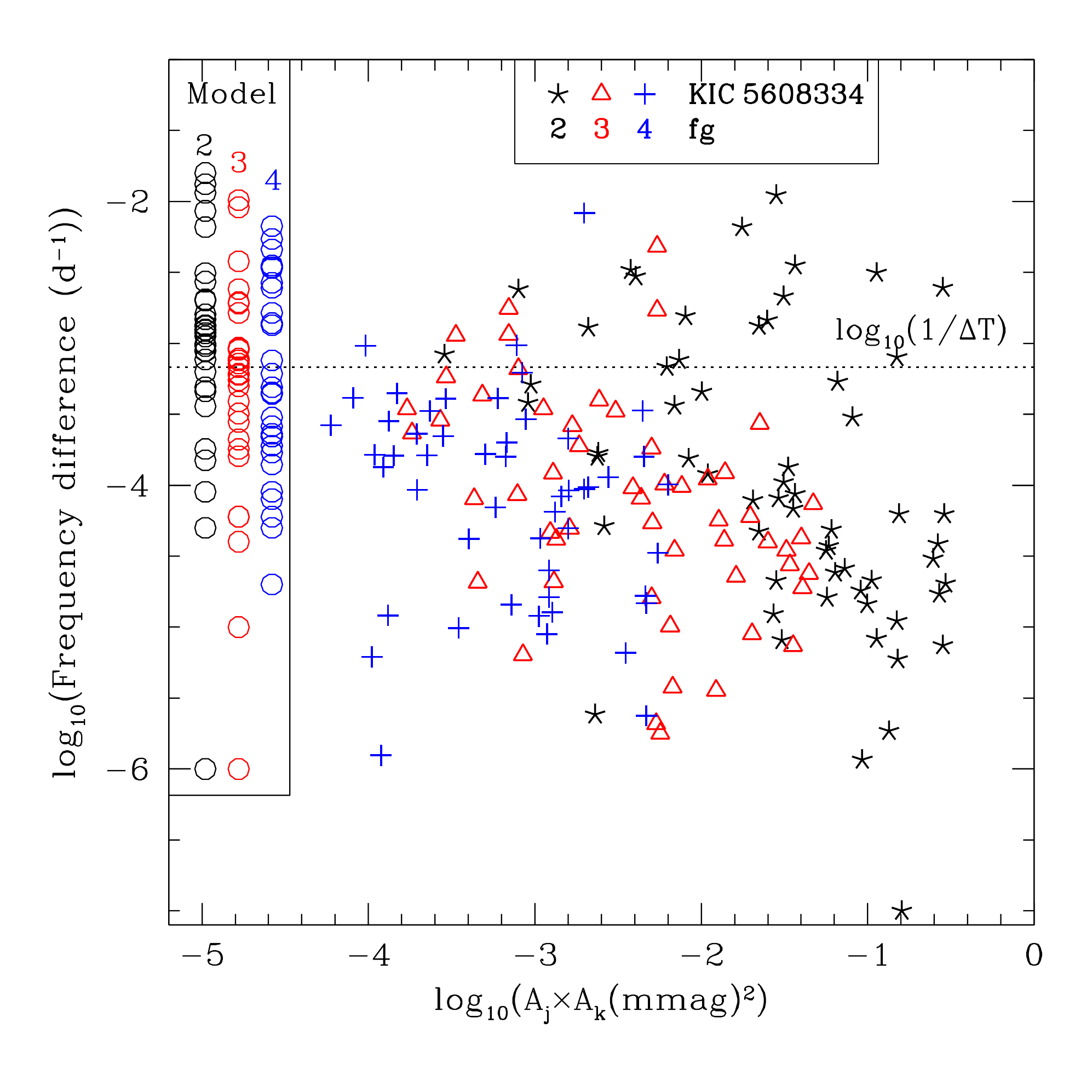}  %{comb_freq_trusted_absdiff_k5608334_m1p6_13_2amp.pdf}
\caption{Frequency difference from the nearest combination frequency ($\nu_j+\nu_k$ or $2\nu_j$)  for every frequency in fg2 (asterisks), fg3 (triangles) and fg4 (plusses) of KIC~5608334 versus the product of the amplitudes for the pair ($j,k$). 
The dotted line indicates the level of $\log_{10}(1/\Delta T)$ with $\Delta T= 1470$~d.
Circles in the inset are for the same frequency difference for linear frequencies of prograde sectoral g~modes in the 1.6-M$_\odot$ model of which parameters are shown in Fig.\,\ref{fig:inert}. (The number near the top of each sequence indicates $|m|$.) 
}
\label{fig:comb_freq}
\end{figure}

Such `frequency lockings' might explain the fact that many  frequencies detected in {\it Kepler} light curves of  KIC~5608334 coincide (within much better than our conservative uncertainty, $1/\Delta T$, see Fig.\ref{fig:comb_freq}) with combination frequencies. These combination frequencies correspond to resonance frequencies, because we identify frequency groups of fg1, $\ldots$, fg4 as prograde sectoral modes of $-m = 1, \ldots, 4$ in a rapidly rotating star.  These identifications are supported by the period spacings of those groups (Fig.\,\ref{fig:inert}). 
In a forthcoming paper, we will discuss more about non-linear effects from a different point of view.

Fig.\,\ref{fig:comb_freq} shows the frequency difference from the nearest combination frequency $(|\nu_i-\nu_j-\nu_k|)$  versus the product of the amplitudes $A_jA_k$ for every frequency $\nu_i$ in the groups fg2, fg3 and fg4. (If mode $i$ belongs to fg2, both $j,k$ should belong to fg1, while if $i$ belongs to fg3, one of $j$ and $k$ should be from fg1 and the other from fg2, while if $i$ belongs to fg4,  both $j,k$ may be from fg2, or $j$ from fg3 and $k$ from fg1, etc.) 
Not all, but many frequencies are very close to combination frequencies, satisfying the three-mode resonance conditions.

Open circles in the inset of Fig.\,\ref{fig:comb_freq} show deviations from the nearest combination frequencies among theoretical linear frequencies for prograde sectoral g~modes. This represents the property of prograde sectoral g~modes discussed in the previous section; i.e., they tend to be nearly in resonance with prograde sectoral g~modes of other azimuthal order $m$. In some cases nearly exact resonance occurs among linear theoretical frequencies (without any non-linear effects), which is consistent with the fact that observed frequencies are sometimes in nearly exact resonance with relatively small non-linear effects (i.e., small $A_jA_k$). This further supports our identification of the observed frequency groups of  KIC~5608334 as prograde sectoral g~modes.

%In addition,  Fig.\,\ref{fig:comb_freq} shows that the vertical range of model (linear) frequencies tends to be higher than those of  KIC~5608334.
Although the extent of the frequency pairs of KIC~5608334 above the dotted line ($1/\Delta T$) in Fig.\,\ref{fig:comb_freq} is comparable to that of model (linear) frequency pairs,  about 85 per cent of the observational points (in contrast to 49 per cent of the theoretical pairs) are located below the dotted line.
This again indicates that pulsation frequencies of KIC~5608334 are modified by non-linear couplings.

\section{Concluding Remarks}

We have identified the four frequency groups fg1, $\ldots$, fg4 of KIC~5608334 as prograde sectoral g~modes with azimuthal orders of $1, 2, 3$ and $4$ strongly influenced by the Coriolis force. At a rotation frequency of $2.2$~d$^{-1}$, those intermediate to high radial order ($\sim20$ to $\sim60$) modes reproduce well the observed frequency range and $\Delta P$-period relation of each frequency group of  KIC~5608334. A comparison of the typical amplitude of each group, using the visibilities for different $\ell$ modes derived for non-rotating models by \citet{das02}, indicates that modes of different $m$ are excited to comparable intrinsic amplitudes and their relative observed amplitudes on the stellar surface are determined by partial (geometric) cancellation.  

With the rotation frequency we can convert observed frequencies in each group to frequencies of the co-rotating frame ($\nu^{\rm co}$). For all frequencies the spin parameters are found to be larger than unity; i.e., $2\nu_{\rm rot}/\nu^{\rm co} > 1$, indicating the importance of the Coriolis force in forming the character of those g~modes. Under such conditions, the frequencies of prograde sectoral modes are approximately proportional to the azimuthal order; i.e., $\nu^{\rm co}\approx |m|\nu_0/n_g$, which indicates formation of frequency groups in the inertial frame, $|m|(\nu_0/n_g+\nu_{\rm rot})$. 
%if $\nu_0/ng < \nu_{\rm rot}$, as in the case of  KIC~5608334 presented in this paper. 
Frequency groups of this type also appear in other rapidly rotating g-mode pulsators, such as Be stars \citep[e.g.][]{wal05,cam08} and Slowly Pulsating B (SPB) stars in young open clusters \citep[e.g.][]{sai17}.  We obtained and discussed for the first time the period spacings in each frequency group confirming the rotational origin of the frequency groups.

Another conspicuous property of the pulsation frequencies of KIC~5608334 is the presence of many frequencies that are nearly or exactly equal to combinations of other frequencies. We discussed the property in relation to the properties of prograde sectoral g~modes under the dominance of Coriolis force, in which frequencies are proportional to $|m|$ even in the co-rotating frame. Then,  the condition of combination frequencies becomes equal to the resonance condition for a non-linear coupling; $\nu^{\rm co}_i \approx \nu^{\rm co}_j + \nu^{\rm co}_k$ with $m_i = m_j + m_k$. 
This explains the presence of many combination frequencies of KIC~5608334.

\section*{Acknowledgements}
We thank Umin Lee for helpful discussions. We also thank Professor John Telting for helpful comments.
This work has made use of data from the European Space Agency (ESA) mission {\it Gaia} (\url{https://www.cosmos.esa.int/gaia}), processed by the {\it Gaia} Data Processing and Analysis Consortium (DPAC, \url{https://www.cosmos.esa.int/web/gaia/dpac/consortium}). Funding for the DPAC has been provided by national institutions, in particular the institutions participating in the {\it Gaia} Multilateral Agreement.  Funding for the Stellar Astrophysics Centre is provided by The Danish National Research Foundation (Grant DNRF106).

\bibliographystyle{mnras}
\bibliography{5608334}

\begin{thebibliography}{}
\makeatletter
\relax
\def\mn@urlcharsother{\let\do\@makeother \do\$\do\&\do\#\do\^\do\_\do\%\do\~}
\def\mn@doi{\begingroup\mn@urlcharsother \@ifnextchar [ {\mn@doi@}
  {\mn@doi@[]}}
\def\mn@doi@[#1]#2{\def\@tempa{#1}\ifx\@tempa\@empty \href
  {http://dx.doi.org/#2} {doi:#2}\else \href {http://dx.doi.org/#2} {#1}\fi
  \endgroup}
\def\mn@eprint#1#2{\mn@eprint@#1:#2::\@nil}
\def\mn@eprint@arXiv#1{\href {http://arxiv.org/abs/#1} {{\tt arXiv:#1}}}
\def\mn@eprint@dblp#1{\href {http://dblp.uni-trier.de/rec/bibtex/#1.xml}
  {dblp:#1}}
\def\mn@eprint@#1:#2:#3:#4\@nil{\def\@tempa {#1}\def\@tempb {#2}\def\@tempc
  {#3}\ifx \@tempc \@empty \let \@tempc \@tempb \let \@tempb \@tempa \fi \ifx
  \@tempb \@empty \def\@tempb {arXiv}\fi \@ifundefined
  {mn@eprint@\@tempb}{\@tempb:\@tempc}{\expandafter \expandafter \csname
  mn@eprint@\@tempb\endcsname \expandafter{\@tempc}}}

\bibitem[\protect\citeauthoryear{{Aerts}, {Christensen-Dalsgaard}  \&
  {Kurtz}}{{Aerts} et~al.}{2010}]{ack10}
{Aerts} C.,  {Christensen-Dalsgaard} J.,   {Kurtz} D.~W.,  2010,
  {Asteroseismology}

\bibitem[\protect\citeauthoryear{{Appourchaux} et~al.,}{{Appourchaux}
  et~al.}{2010}]{app10}
{Appourchaux} T.,  et~al., 2010, \mn@doi [\aapr] {10.1007/s00159-009-0027-z},
  \href {http://adsabs.harvard.edu/abs/2010A%26ARv..18..197A} {18, 197}

\bibitem[\protect\citeauthoryear{{Ballot}, {Ligni{\`e}res}, {Prat}, {Reese}  \&
  {Rieutord}}{{Ballot} et~al.}{2012}]{bal12}
{Ballot} J.,  {Ligni{\`e}res} F.,  {Prat} V.,  {Reese} D.~R.,   {Rieutord} M.,
  2012, in {Shibahashi} H.,  {Takata} M.,   {Lynas-Gray} A.~E.,  eds,
  Astronomical Society of the Pacific Conference Series Vol. 462, Progress in
  Solar/Stellar Physics with Helio- and Asteroseismology. p.~389 (\mn@eprint
  {arXiv} {1109.6856})

\bibitem[\protect\citeauthoryear{{Balona}, {Krisciunas}  \& {Cousins}}{{Balona}
  et~al.}{1994}]{balona1994}
{Balona} L.~A.,  {Krisciunas} K.,   {Cousins} A.~W.~J.,  1994, \mn@doi [\mnras]
  {10.1093/mnras/270.4.905}, \href
  {http://adsabs.harvard.edu/abs/1994MNRAS.270..905B} {270, 905}

\bibitem[\protect\citeauthoryear{{Berthomieu}, {Gonczi}, {Graff}, {Provost}  \&
  {Rocca}}{{Berthomieu} et~al.}{1978}]{ber78}
{Berthomieu} G.,  {Gonczi} G.,  {Graff} P.,  {Provost} J.,   {Rocca} A.,  1978,
  \aap, \href {http://ads.nao.ac.jp/abs/1978A%26A....70..597B} {70, 597}

\bibitem[\protect\citeauthoryear{{Borucki} et~al.,}{{Borucki}
  et~al.}{2010}]{bor10}
{Borucki} W.~J.,  et~al., 2010, \mn@doi [Science] {10.1126/science.1185402},
  \href {http://adsabs.harvard.edu/abs/2010Sci...327..977B} {327, 977}

\bibitem[\protect\citeauthoryear{{Bouabid}, {Dupret}, {Salmon},
  {Montalb{\'a}n}, {Miglio}  \& {Noels}}{{Bouabid} et~al.}{2013}]{bou13}
{Bouabid} M.-P.,  {Dupret} M.-A.,  {Salmon} S.,  {Montalb{\'a}n} J.,  {Miglio}
  A.,   {Noels} A.,  2013, \mn@doi [\mnras] {10.1093/mnras/sts517}, \href
  {http://ads.nao.ac.jp/abs/2013MNRAS.429.2500B} {429, 2500}

\bibitem[\protect\citeauthoryear{{Cameron} et~al.,}{{Cameron}
  et~al.}{2008}]{cam08}
{Cameron} C.,  et~al., 2008, \mn@doi [\apj] {10.1086/590369}, \href
  {http://adsabs.harvard.edu/abs/2008ApJ...685..489C} {685, 489}

\bibitem[\protect\citeauthoryear{{Cousins}}{{Cousins}}{1924}]{cousins1924}
{Cousins} A.~W.~J.,  1924, \mn@doi [\mnras] {10.1093/mnras/84.8.620}, \href
  {http://adsabs.harvard.edu/abs/1924MNRAS..84..620C} {84, 620}

\bibitem[\protect\citeauthoryear{{Cousins}}{{Cousins}}{1992}]{cousins1992}
{Cousins} A.~W.~J.,  1992, The Observatory, \href
  {http://adsabs.harvard.edu/abs/1992Obs...112...53C} {112, 53}

\bibitem[\protect\citeauthoryear{{Cousins}}{{Cousins}}{1994}]{cousins1994}
{Cousins} A.~W.~J.,  1994, The Observatory, \href
  {http://adsabs.harvard.edu/abs/1994Obs...114...51C} {114, 51}

\bibitem[\protect\citeauthoryear{{Cousins} \& {Caldwell}}{{Cousins} \&
  {Caldwell}}{2001}]{cousins2001}
{Cousins} A.~W.~J.,  {Caldwell} J.~A.~R.,  2001, \mn@doi [\mnras]
  {10.1046/j.1365-8711.2001.04187.x}, \href
  {http://adsabs.harvard.edu/abs/2001MNRAS.323..380C} {323, 380}

\bibitem[\protect\citeauthoryear{{Cousins} \& {Warren}}{{Cousins} \&
  {Warren}}{1963}]{cousins1963}
{Cousins} A.~W.~J.,  {Warren} P.~R.,  1963, Monthly Notes of the Astronomical
  Society of South Africa, \href
  {http://adsabs.harvard.edu/abs/1963MNSSA..22...65C} {22, 65}

\bibitem[\protect\citeauthoryear{{Cousins}, {Caldwell}  \& {Menzies}}{{Cousins}
  et~al.}{1989}]{cousins1989}
{Cousins} A.~W.~J.,  {Caldwell} J.~A.~R.,   {Menzies} J.~W.,  1989, Information
  Bulletin on Variable Stars, \href
  {http://adsabs.harvard.edu/abs/1989IBVS.3412....1C} {3412}

\bibitem[\protect\citeauthoryear{{Daszy{\'n}ska-Daszkiewicz}, {Dziembowski},
  {Pamyatnykh}  \& {Goupil}}{{Daszy{\'n}ska-Daszkiewicz} et~al.}{2002}]{das02}
{Daszy{\'n}ska-Daszkiewicz} J.,  {Dziembowski} W.~A.,  {Pamyatnykh} A.~A.,
  {Goupil} M.-J.,  2002, \mn@doi [\aap] {10.1051/0004-6361:20020911}, \href
  {http://adsabs.harvard.edu/abs/2002A%26A...392..151D} {392, 151}

\bibitem[\protect\citeauthoryear{{Dupret}, {Grigahc{\`e}ne}, {Garrido},
  {Gabriel}  \& {Scuflaire}}{{Dupret} et~al.}{2005}]{dup05}
{Dupret} M.-A.,  {Grigahc{\`e}ne} A.,  {Garrido} R.,  {Gabriel} M.,
  {Scuflaire} R.,  2005, \mn@doi [\aap] {10.1051/0004-6361:20041817}, \href
  {http://adsabs.harvard.edu/abs/2005A%26A...435..927D} {435, 927}

\bibitem[\protect\citeauthoryear{{Dziembowski}}{{Dziembowski}}{1982}]{dzi82}
{Dziembowski} W.,  1982, \actaa, \href
  {http://adsabs.harvard.edu/abs/1982AcA....32..147D} {32, 147}

\bibitem[\protect\citeauthoryear{{Flower}}{{Flower}}{1996}]{flo96}
{Flower} P.~J.,  1996, \mn@doi [\apj] {10.1086/177785}, \href
  {http://adsabs.harvard.edu/abs/1996ApJ...469..355F} {469, 355}

\bibitem[\protect\citeauthoryear{{Fossat} et~al.,}{{Fossat}
  et~al.}{2017}]{fos17}
{Fossat} E.,  et~al., 2017, \mn@doi [\aap] {10.1051/0004-6361/201730460}, \href
  {http://adsabs.harvard.edu/abs/2017A%26A...604A..40F} {604, A40}

\bibitem[\protect\citeauthoryear{{Gaia Collaboration} et~al.,}{{Gaia
  Collaboration} et~al.}{2016}]{gaia16}
{Gaia Collaboration} et~al., 2016, \mn@doi [\aap]
  {10.1051/0004-6361/201629512}, \href
  {http://adsabs.harvard.edu/abs/2016A%26A...595A...2G} {595, A2}

\bibitem[\protect\citeauthoryear{Handberg}{Handberg}{2017}]{timeseries_tools}
Handberg R.,  2017, rhandberg/timeseries: Initial release,
  \mn@doi{10.5281/zenodo.400605}, \url {https://doi.org/10.5281/zenodo.400605}

\bibitem[\protect\citeauthoryear{{Iglesias} \& {Rogers}}{{Iglesias} \&
  {Rogers}}{1996}]{igl96}
{Iglesias} C.~A.,  {Rogers} F.~J.,  1996, \mn@doi [\apj] {10.1086/177381},
  \href {http://adsabs.harvard.edu/abs/1996ApJ...464..943I} {464, 943}

\bibitem[\protect\citeauthoryear{{Kilkenny}}{{Kilkenny}}{2001}]{kilkenny2001}
{Kilkenny} D.,  2001, The Observatory, \href
  {http://adsabs.harvard.edu/abs/2001Obs...121..350K} {121, 350}

\bibitem[\protect\citeauthoryear{{Kurtz}, {Saio}, {Takata}, {Shibahashi},
  {Murphy}  \& {Sekii}}{{Kurtz} et~al.}{2014}]{kur14}
{Kurtz} D.~W.,  {Saio} H.,  {Takata} M.,  {Shibahashi} H.,  {Murphy} S.~J.,
  {Sekii} T.,  2014, \mn@doi [\mnras] {10.1093/mnras/stu1329}, \href
  {http://adsabs.harvard.edu/abs/2014MNRAS.444..102K} {444, 102}

\bibitem[\protect\citeauthoryear{{Kurtz}, {Shibahashi}, {Murphy}, {Bedding}  \&
  {Bowman}}{{Kurtz} et~al.}{2015}]{kur15}
{Kurtz} D.~W.,  {Shibahashi} H.,  {Murphy} S.~J.,  {Bedding} T.~R.,   {Bowman}
  D.~M.,  2015, \mn@doi [\mnras] {10.1093/mnras/stv868}, \href
  {http://adsabs.harvard.edu/abs/2015MNRAS.450.3015K} {450, 3015}

\bibitem[\protect\citeauthoryear{{Lee} \& {Baraffe}}{{Lee} \&
  {Baraffe}}{1995}]{lee95}
{Lee} U.,  {Baraffe} I.,  1995, \aap, \href
  {http://adsabs.harvard.edu/abs/1995A%26A...301..419L} {301, 419}

\bibitem[\protect\citeauthoryear{{Lee} \& {Saio}}{{Lee} \&
  {Saio}}{1987}]{lee87b}
{Lee} U.,  {Saio} H.,  1987, \mn@doi [\mnras] {10.1093/mnras/224.3.513}, \href
  {http://adsabs.harvard.edu/abs/1987MNRAS.224..513L} {224, 513}

\bibitem[\protect\citeauthoryear{{Lee} \& {Saio}}{{Lee} \&
  {Saio}}{1997}]{lee97}
{Lee} U.,  {Saio} H.,  1997, \apj, \href
  {http://adsabs.harvard.edu/abs/1997ApJ...491..839L} {491, 839}

\bibitem[\protect\citeauthoryear{{Lenz} \& {Breger}}{{Lenz} \&
  {Breger}}{2005}]{period04}
{Lenz} P.,  {Breger} M.,  2005, \mn@doi [Communications in Asteroseismology]
  {10.1553/cia146s53}, \href
  {http://adsabs.harvard.edu/abs/2005CoAst.146...53L} {146, 53}

\bibitem[\protect\citeauthoryear{{McNamara}, {Jackiewicz}  \&
  {McKeever}}{{McNamara} et~al.}{2012}]{mcnamara2012}
{McNamara} B.~J.,  {Jackiewicz} J.,   {McKeever} J.,  2012, \mn@doi [\aj]
  {10.1088/0004-6256/143/4/101}, \href
  {http://adsabs.harvard.edu/abs/2012AJ....143..101M} {143, 101}

\bibitem[\protect\citeauthoryear{{Miglio}, {Montalb{\'a}n}, {Noels}  \&
  {Eggenberger}}{{Miglio} et~al.}{2008}]{mig08}
{Miglio} A.,  {Montalb{\'a}n} J.,  {Noels} A.,   {Eggenberger} P.,  2008,
  \mn@doi [\mnras] {10.1111/j.1365-2966.2008.13112.x}, \href
  {http://adsabs.harvard.edu/abs/2008MNRAS.386.1487M} {386, 1487}

\bibitem[\protect\citeauthoryear{{Murphy}, {Bedding}, {Niemczura}, {Kurtz}  \&
  {Smalley}}{{Murphy} et~al.}{2015}]{mur15}
{Murphy} S.~J.,  {Bedding} T.~R.,  {Niemczura} E.,  {Kurtz} D.~W.,   {Smalley}
  B.,  2015, \mn@doi [\mnras] {10.1093/mnras/stu2749}, \href
  {http://adsabs.harvard.edu/abs/2015MNRAS.447.3948M} {447, 3948}

\bibitem[\protect\citeauthoryear{{Murphy}, {Fossati}, {Bedding}, {Saio},
  {Kurtz}, {Grassitelli}  \& {Wang}}{{Murphy} et~al.}{2016}]{mur16}
{Murphy} S.~J.,  {Fossati} L.,  {Bedding} T.~R.,  {Saio} H.,  {Kurtz} D.~W.,
  {Grassitelli} L.,   {Wang} E.~S.,  2016, \mn@doi [\mnras]
  {10.1093/mnras/stw705}, \href
  {http://adsabs.harvard.edu/abs/2016MNRAS.459.1201M} {459, 1201}

\bibitem[\protect\citeauthoryear{{Niemczura} et~al.,}{{Niemczura}
  et~al.}{2015}]{nie15}
{Niemczura} E.,  et~al., 2015, \mn@doi [\mnras] {10.1093/mnras/stv528}, \href
  {http://adsabs.harvard.edu/abs/2015MNRAS.450.2764N} {450, 2764}

\bibitem[\protect\citeauthoryear{{Ouazzani}, {Salmon}, {Antoci}, {Bedding},
  {Murphy}  \& {Roxburgh}}{{Ouazzani} et~al.}{2017}]{oua17}
{Ouazzani} R.-M.,  {Salmon} S.~J.~A.~J.,  {Antoci} V.,  {Bedding} T.~R.,
  {Murphy} S.~J.,   {Roxburgh} I.~W.,  2017, \mn@doi [\mnras]
  {10.1093/mnras/stw2717}, \href
  {http://adsabs.harvard.edu/abs/2017MNRAS.465.2294O} {465, 2294}

\bibitem[\protect\citeauthoryear{{P{\'a}pics} et~al.,}{{P{\'a}pics}
  et~al.}{2017}]{pap17}
{P{\'a}pics} P.~I.,  et~al., 2017, \mn@doi [\aap]
  {10.1051/0004-6361/201629814}, \href
  {http://adsabs.harvard.edu/abs/2017A%26A...598A..74P} {598, A74}

\bibitem[\protect\citeauthoryear{{Paxton} et~al.,}{{Paxton}
  et~al.}{2013}]{pax13}
{Paxton} B.,  et~al., 2013, \mn@doi [\apjs] {10.1088/0067-0049/208/1/4}, \href
  {http://adsabs.harvard.edu/abs/2013ApJS..208....4P} {208, 4}

\bibitem[\protect\citeauthoryear{{Reese}, {MacGregor}, {Jackson}, {Skumanich}
  \& {Metcalfe}}{{Reese} et~al.}{2009}]{ree09}
{Reese} D.~R.,  {MacGregor} K.~B.,  {Jackson} S.,  {Skumanich} A.,   {Metcalfe}
  T.~S.,  2009, \mn@doi [\aap] {10.1051/0004-6361/200811510}, \href
  {http://adsabs.harvard.edu/abs/2009A%26A...506..189R} {506, 189}

\bibitem[\protect\citeauthoryear{{Saio}, {Kurtz}, {Takata}, {Shibahashi},
  {Murphy}, {Sekii}  \& {Bedding}}{{Saio} et~al.}{2015}]{sai15}
{Saio} H.,  {Kurtz} D.~W.,  {Takata} M.,  {Shibahashi} H.,  {Murphy} S.~J.,
  {Sekii} T.,   {Bedding} T.~R.,  2015, \mn@doi [\mnras]
  {10.1093/mnras/stu2696}, \href
  {http://adsabs.harvard.edu/abs/2015MNRAS.447.3264S} {447, 3264}

\bibitem[\protect\citeauthoryear{{Saio}, {Ekstr{\"o}m}, {Mowlavi}, {Georgy},
  {Saesen}, {Eggenberger}, {Semaan}  \& {Salmon}}{{Saio} et~al.}{2017}]{sai17}
{Saio} H.,  {Ekstr{\"o}m} S.,  {Mowlavi} N.,  {Georgy} C.,  {Saesen} S.,
  {Eggenberger} P.,  {Semaan} T.,   {Salmon} S.~J.~A.~J.,  2017, \mn@doi
  [\mnras] {10.1093/mnras/stx346}, \href
  {http://adsabs.harvard.edu/abs/2017MNRAS.467.3864S} {467, 3864}

\bibitem[\protect\citeauthoryear{{Saio}, {Kurtz}, {Murphy}, {Antoci}  \&
  {Lee}}{{Saio} et~al.}{2018}]{sai18}
{Saio} H.,  {Kurtz} D.~W.,  {Murphy} S.~J.,  {Antoci} V.~L.,   {Lee} U.,  2018,
  \mn@doi [\mnras] {10.1093/mnras/stx2962}, \href
  {http://adsabs.harvard.edu/abs/2018MNRAS.474.2774S} {474, 2774}

\bibitem[\protect\citeauthoryear{{Scargle}}{{Scargle}}{1982}]{sca82}
{Scargle} J.~D.,  1982, \mn@doi [\apj] {10.1086/160554}, \href
  {http://adsabs.harvard.edu/abs/1982ApJ...263..835S} {263, 835}

\bibitem[\protect\citeauthoryear{{Schmid} et~al.,}{{Schmid}
  et~al.}{2015}]{schmid2015}
{Schmid} V.~S.,  et~al., 2015, \mn@doi [\aap] {10.1051/0004-6361/201526945},
  \href {http://adsabs.harvard.edu/abs/2015A%26A...584A..35S} {584, A35}

\bibitem[\protect\citeauthoryear{{Stobie}}{{Stobie}}{1971}]{stobie1971}
{Stobie} R.~S.,  1971, Monthly Notes of the Astronomical Society of South
  Africa, \href {http://adsabs.harvard.edu/abs/1971MNSSA..30...31S} {30, 31}

\bibitem[\protect\citeauthoryear{{Unno}, {Osaki}, {Ando}, {Saio}  \&
  {Shibahashi}}{{Unno} et~al.}{1989}]{unno}
{Unno} W.,  {Osaki} Y.,  {Ando} H.,  {Saio} H.,   {Shibahashi} H.,  1989,
  {Nonradial oscillations of stars}

\bibitem[\protect\citeauthoryear{{Van Reeth} et~al.,}{{Van Reeth}
  et~al.}{2015a}]{vanreeth2015}
{Van Reeth} T.,  et~al., 2015a, \mn@doi [\apjs] {10.1088/0067-0049/218/2/27},
  \href {http://adsabs.harvard.edu/abs/2015ApJS..218...27V} {218, 27}

\bibitem[\protect\citeauthoryear{{Van Reeth} et~al.,}{{Van Reeth}
  et~al.}{2015b}]{vanr15}
{Van Reeth} T.,  et~al., 2015b, \mn@doi [\aap] {10.1051/0004-6361/201424585},
  \href {http://adsabs.harvard.edu/abs/2015A%26A...574A..17V} {574, A17}

\bibitem[\protect\citeauthoryear{{Van Reeth}, {Tkachenko}  \& {Aerts}}{{Van
  Reeth} et~al.}{2016}]{vanr16}
{Van Reeth} T.,  {Tkachenko} A.,   {Aerts} C.,  2016, \mn@doi [\aap]
  {10.1051/0004-6361/201628616}, \href
  {http://adsabs.harvard.edu/abs/2016A%26A...593A.120V} {593, A120}

\bibitem[\protect\citeauthoryear{{Walker} et~al.,}{{Walker}
  et~al.}{2005}]{wal05}
{Walker} G.~A.~H.,  et~al., 2005, \mn@doi [\apjl] {10.1086/499362}, \href
  {http://adsabs.harvard.edu/abs/2005ApJ...635L..77W} {635, L77}

\makeatother
\end{thebibliography}

\appendix

\section{Amplitude distribution of g modes on the stellar surface }
\begin{figure*}
\includegraphics[width=0.48\columnwidth]{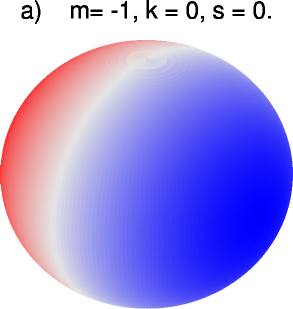}   %{m-1k0_spn0_6cm_a.pdf}
\hspace{0.02\columnwidth}
%\hspace{0.04\columnwidth}
\includegraphics[width=0.48\columnwidth]{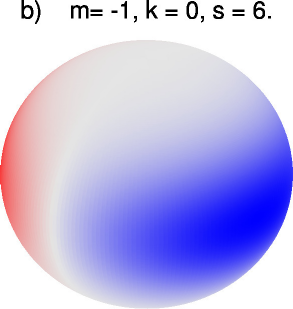}   %{m-1k0_spn6_6cm_b.pdf}
\hspace{0.02\columnwidth}
%\vspace{0.5cm}
\includegraphics[width=0.48\columnwidth]{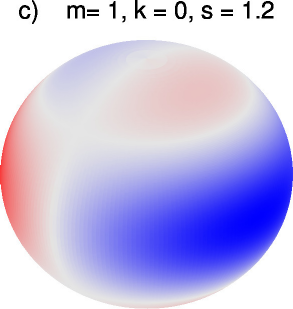}   %{m1k0_spn1p2_6cm_c.pdf}
\hspace{0.02\columnwidth}
\includegraphics[width=0.48\columnwidth]{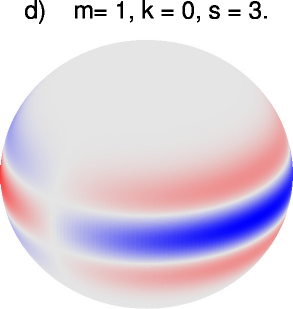}   %{m1k0_spn3_6cm_d.pdf}

\vspace{0.5cm}
\includegraphics[width=0.48\columnwidth]{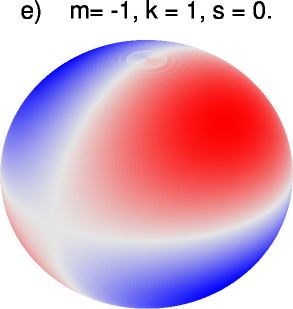}   %{m-1k1_spn0_6cm_e.pdf}
\hspace{0.02\columnwidth}
\includegraphics[width=0.48\columnwidth]{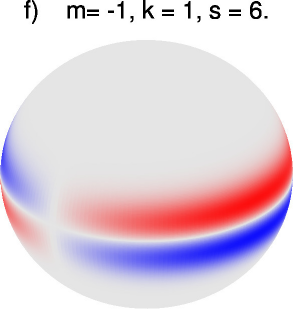}    %{m-1k1_spn6_6cm_f.pdf}
\hspace{0.02\columnwidth}
%\vspace{0.5cm}
\includegraphics[width=0.47\columnwidth]{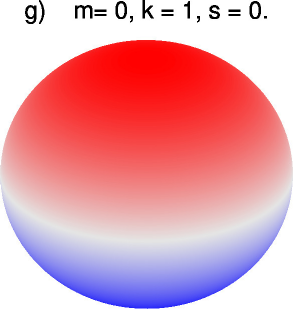}   %{m0k1_spn0_6cm_g.pdf}
\hspace{0.02\columnwidth}
\includegraphics[width=0.47\columnwidth]{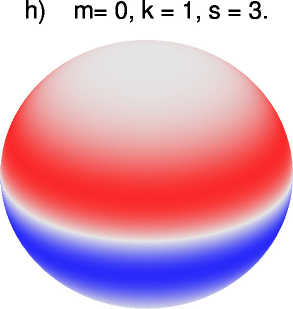}   %{m0k1_spn3_6cm_h.pdf}

\caption{Some of the amplitude distributions of temperature variations (or radial displacements) of $m=0,\pm1$ g modes on the stellar surface, with and without rotation. (In this paper, we adopt the convention that $m <0$ for prograde and $m > 0$ for retrograde modes.)
An inclination angle of $60^\circ$ is adopted.
 The rotation affects the amplitude distribution of a low-frequency mode through the spin parameter,   $s\equiv2\nu_{\rm rot}/\nu^{\rm co}$, with $\nu_{\rm rot}$ and $\nu^{\rm co}$ being the rotation frequency and the pulsation frequency in the co-rotating frame, respectively. The parameter, $k$ ($>0$ for g modes)  \citep[adopted from][]{lee97}, specifies the parity and the order of latitudinal amplitude distribution.  In the non-rotating case the latitudinal degree $\ell$ is given as $\ell = |m| + k$;   $k=0$ means the first even mode (symmetric to the equator), while $k=1$ the first tesseral (odd) mode. For slow rotation ($s < 1$),  $k=0$ modes are sectoral modes (no latitudinal nodal line).  If $s > 1$, however,  retrograde $k=0$ modes have two latitudinal nodal lines (one in each hemisphere), while prograde $k=0$ modes remain sectoral modes.     
}
\label{fig:patterns}
\end{figure*}

The amplitude distribution of a nonradial pulsation mode on the stellar surface is described by a spherical harmonic $Y_\ell^m(\theta,\phi)$ in a nonrotating star. The distribution is modified in a rotating star because of the effect of the Coriolis force. This effect is significant if the spin parameter $s\equiv 2\nu_{\rm rot}/\nu^{\rm co}$ is greater than unity, where $\nu_{\rm rot}$ and $\nu^{\rm co}$ are rotation frequency and the pulsation (g- mode) frequency of a pulsation mode in the co-rotating frame, respectively. Fig.~\ref{fig:patterns} shows some examples, where the angular dependences of g modes are ordered by $m$ and $k$ \citep[adopting from][]{lee97}; $\ell = |m| + k$ at $s=0$ (we use in this paper negative $m(<0)$ for prograde modes).    

Panels a) and b) of Fig.~\ref{fig:patterns} are for prograde sectoral modes at $s=0$ and $s=6$, respectively. The prograde sectoral modes remain sectoral even in a rapidly rotating star. However,  retrograde g modes differ significantly, as shown in panels c) and d). Although a retrograde $k=0$ mode keeps the sectoral character if $s < 1$, two latitudinal nodal  lines appear for $s > 1$ (i.e., no longer sectoral) and the amplitude become strongly confined to an equatorial zone as the spin parameter $s$ increases.

Panels e) and f) are for a prograde  tesseral mode ($m=-1, k=1$) at $s=0$ ($\ell=2$) and at $s=3$, respectively.
Tesseral modes also get strongly confined to an equatorial zone if $s > 1$. 

Finally, panels  g) and h) are for a  zonal ($m=0$) mode of $k=1$ at $s=0$ ($\ell=1$) and $s=3$, respectively. Again, the  amplitude of a zonal mode tends to be concentrated toward the equator.

Thus, in a relatively rapidly rotating star, prograde sectoral modes ($m<0,k=0$) are most visible among g modes. This explains why we detect prograde sectoral modes in KIC~5608334 and why prograde sectoral g modes are predominantly detected in moderately to rapidly rotating $\gamma$ Dor stars \citep[e.g. ][]{vanr16} and SPB stars \citep[e.g. ][]{pap17}. 

\begin{figure*}
\includegraphics[width=0.48\columnwidth]{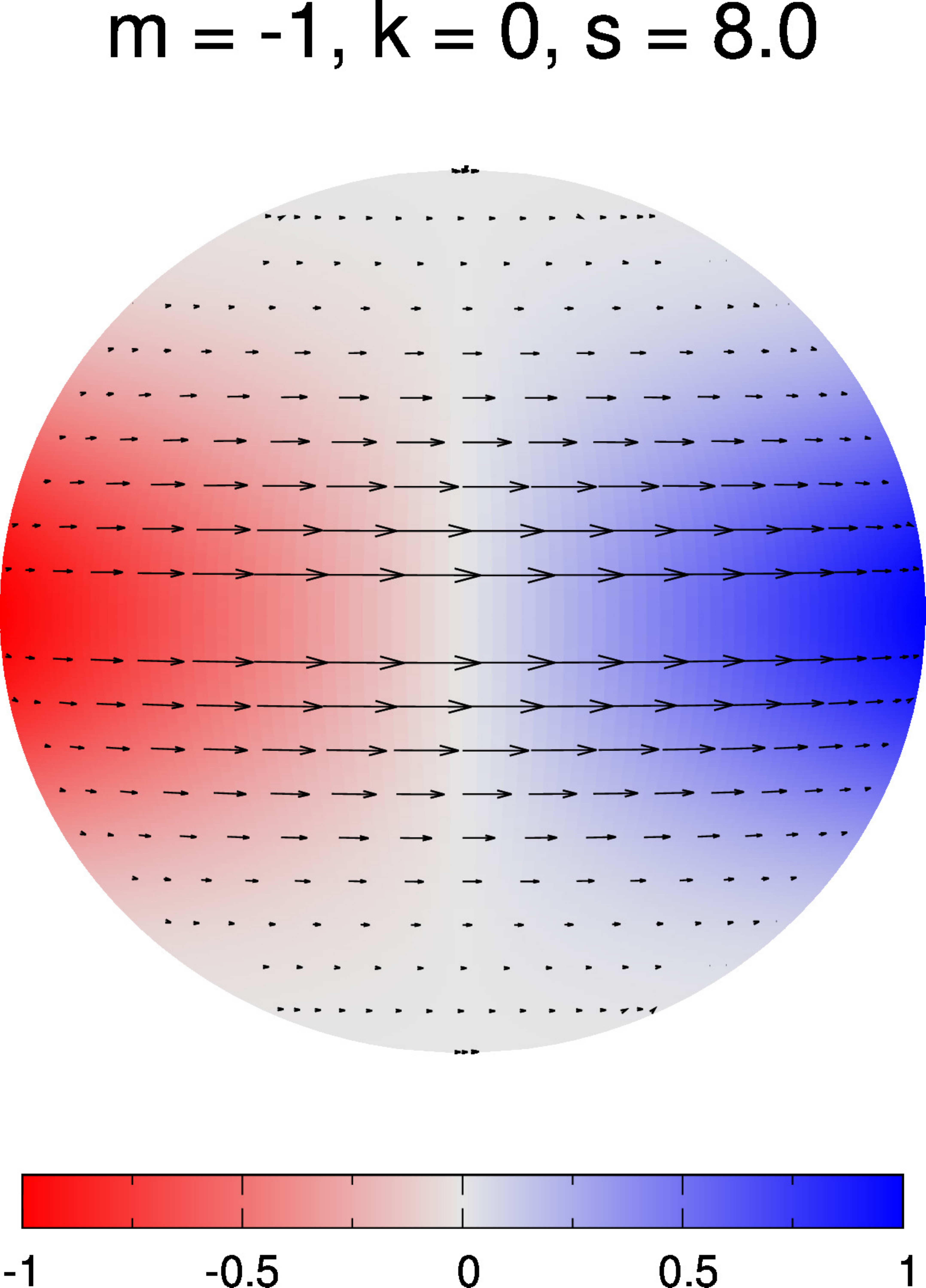} 
\hspace{0.02\columnwidth}
\includegraphics[width=0.48\columnwidth]{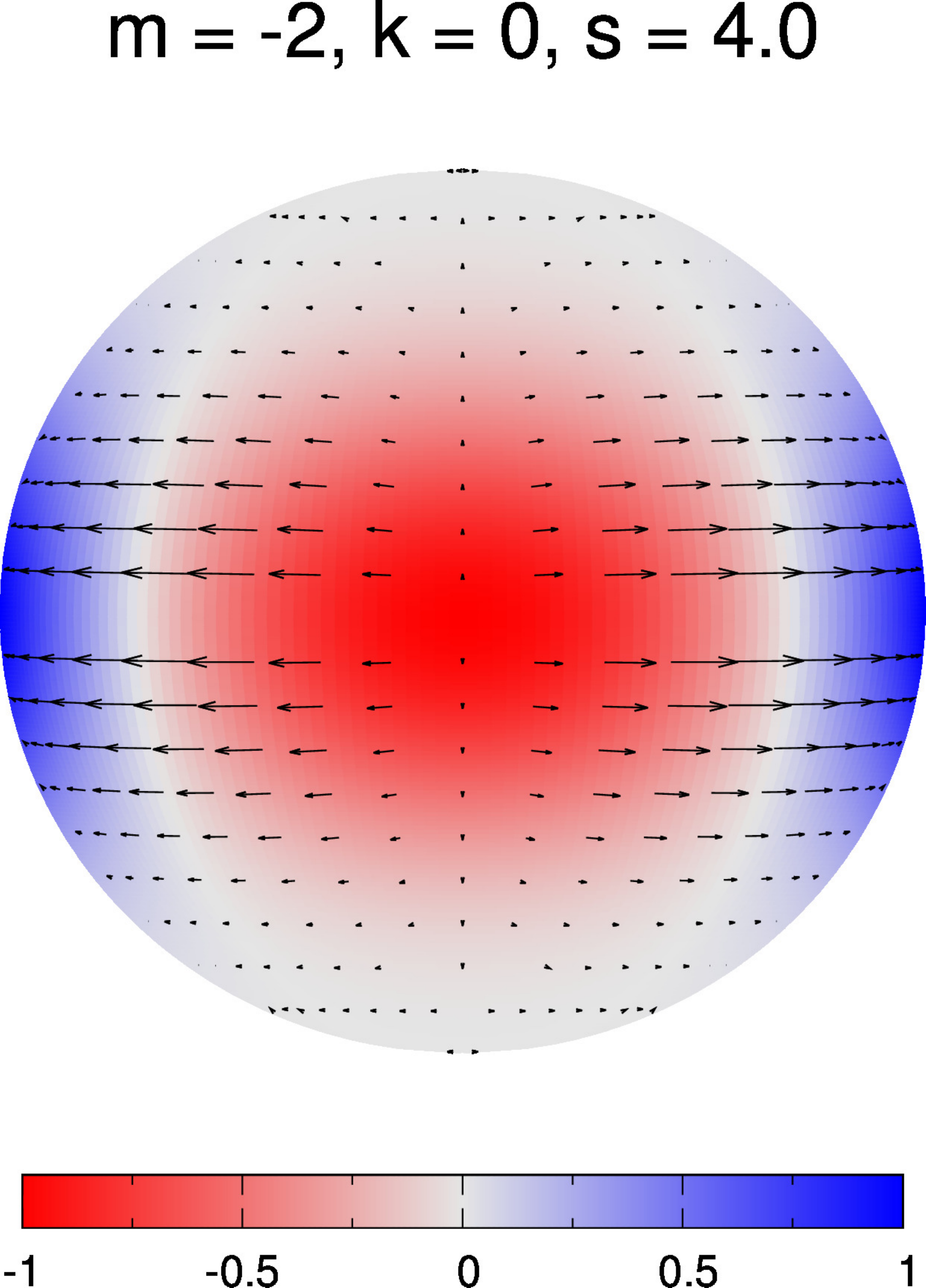} 
\hspace{0.02\columnwidth}
\includegraphics[width=0.48\columnwidth]{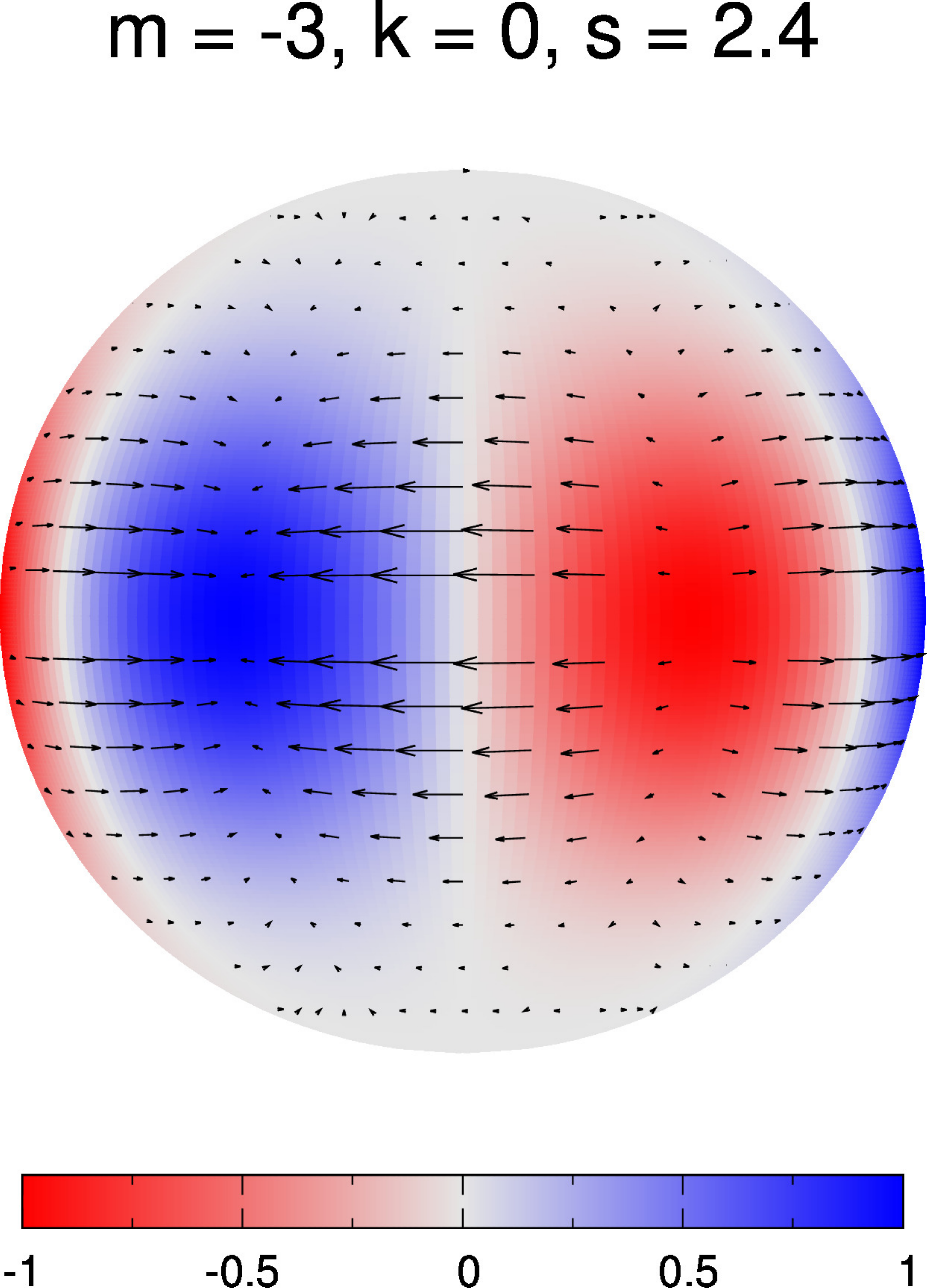} 
\hspace{0.02\columnwidth}
\includegraphics[width=0.48\columnwidth]{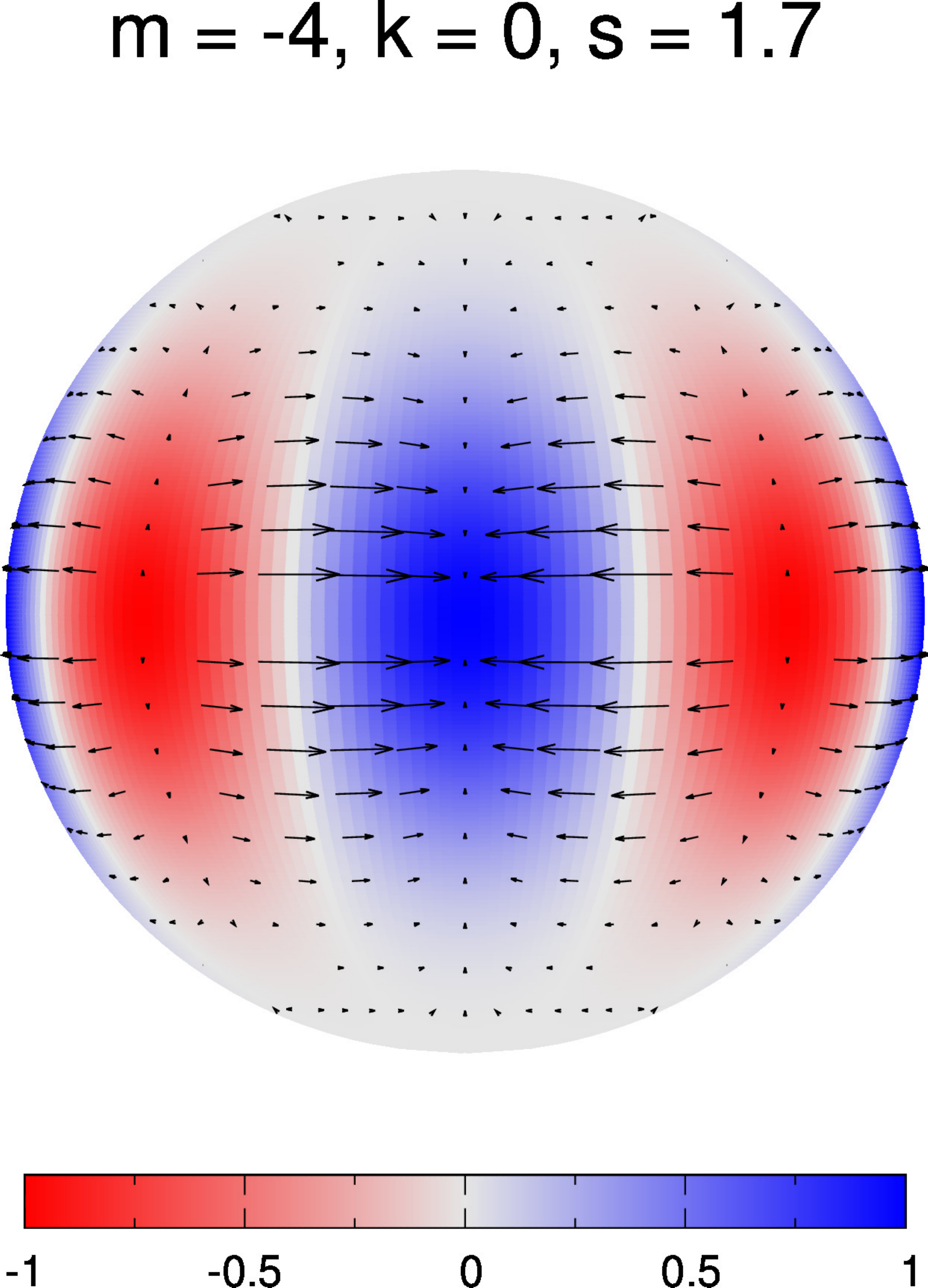}
\caption{Distributions of temperature variations (or radial displacements; colour coded) and horizontal displacements (arrows) predicted for typical g mode pulsations in the frequency groups of fg1 ($m=-1$), fg2 ($m=-2$), fg3 ($m=-3$), and fg4 ($m=-4$) of KIC~5608334. The inclination angle is $90^\circ$. The value of spin parameter ($s$) adopted for each case corresponds to a middle frequency of each frequency group. Horizontal displacements of g modes with large spin parameters are predominantly azimuthal.}
\label{fig:A2} 
\end{figure*}

Fig.~\ref{fig:A2} shows amplitude distributions of the temperature variations (or radial displacement; colour coded) and horizontal displacements (arrows) for typical g mode pulsations in the frequency groups of KIC~5608334 (see Fig.~\ref{fig:ft}). We have identified the groups fg1, fg2, fg3, and fg4 as prograde sectoral g modes of $-m=1$, 2, 3, and 4, respectively. The spin parameter $s$ adopted for each case in this figure corresponds to the middle frequency of each frequency group and the rotation frequency $2.20$~d$^{-1}$. The spin parameters are largest for g modes in fg1 and smallest for those of fg4, although they are still larger than unity. As Fig.\,\ref{fig:A2} indicates, the horizontal displacements are nearly azimuthal in g mode pulsations  with large spin parameters.

\newpage

 \begin{table}
  \caption{Frequency list for KIC 5608334}
  \label{tab:freq}
  \begin{tabular}{cccccc}
    \hline
%    $f_i$ & Frequency & Frequency error & Amplitude & Amplitude error & parent modes \\
    $f_i$ & Frequency  & Amplitude  & parent modes \\
%     & [c/d] & [c/d] & [ppm] & [ppm] &  \\
      & [d$^{-1}$]  & [ppm]  &  \\
   \hline
$\nu_{1}$	&	$2.862775	\pm	0.000002$	&	$541.9	\pm	2.9$	&		\\
$\nu_{2}$	&	$2.882291	\pm	0.000002$	&	$532.1	\pm	3.1$	&		\\
$\nu_{3}$	&	$2.795446	\pm	0.000002$	&	$497.2 \pm 2.5$	&		\\
$\nu_{4}$	&	$2.925108	\pm	0.000004$	&	$301.3	\pm	3.5$	&		\\
$\nu_{5}$	&	$2.781019 \pm	0.000003$	&	$284.6	\pm	2.6$	&	 $3\nu_{1} -\nu_{2} -\nu_{4}$ 	 \\
$\nu_{6}$	&	$3.001873	\pm	0.000005$	&	$269.7	\pm	3.9$	&		\\
$\nu_{7}$	&	$3.179237	\pm	0.000002$	&	$212.4	\pm	1.3$	&	 $-2\nu_{1} -\nu_{3}+4\nu_{4}$ 	 \\
$\nu_{8}$	&	$2.974464	\pm	0.000008$	&	$186.6	\pm	4.1$	&		\\
$\nu_{9}$	&	$3.099039	\pm	0.000005$	&	$132.6	\pm	1.8$	&	    $2\nu_{2}  -2\nu_{3}+  \nu_{4}$ 	 \\
$\nu_{10}$	&	$2.693984 \pm	0.000008$	&	$121.7	\pm	2.5$	&		\\
$\nu_{11}$	&	$2.665894	\pm	0.000008$	&	$117.5	\pm	2.6$	&	       $2\nu_{3}  -\nu_{4}$ 	 \\
$\nu_{12}$	&	$3.064033	\pm	0.000008$	&	$107.2	\pm	2.3$	&		\\
$\nu_{13}$	&	$5.720543	\pm	0.000006$	&	$87.0	\pm	1.4$	&	       $\nu_{3}+\nu_{4}$ 	 \\
$\nu_{14}$	&	$5.764574	\pm	0.000006$	&	$83.4	\pm	1.3$	&	    $2\nu_{2}$ 	 \\
$\nu_{15	}$	&	$5.479216	\pm	0.000006$	&	$74.9	\pm	1.2$	&		\\
$\nu_{16}$	&	$5.807399	\pm	0.000007$	&	$73.9	\pm	1.3$	&	    $\nu_{2} + \nu_{4}$ 	 \\
$\nu_{17}$	&	$5.725531	\pm	0.000007$	&	$73.6	\pm	1.3$	&	 $2\nu_{1}$ 	 \\
$\nu_{18	}$	&	$3.031610	\pm	0.000017$	&	$68.9	\pm	3.2$	&		\\
$\nu_{19}$	&	$5.590923	\pm	0.000007$	&	$67.1	\pm	1.3$	&	       $2\nu_{3}$ 	 \\
$\nu_{20}$	&	$2.734104	\pm	0.000014$	&	$67.0	\pm	2.4$	&	 $\nu_{1}+ 3\nu_{2}  -3\nu_{4}$ 	 \\
$\nu_{21	}$	&	$6.061520	\pm	0.000005$	&	$65.4	\pm	0.9$	&		\\
$\nu_{22}$	&	$5.619004	\pm	0.000008$	&	$63.0	\pm	1.3$	&	 $3\nu_{1}  -2\nu_{2}+ \nu_{3}$ 	 \\
$\nu_{23	}$	&	$5.946340	\pm	0.000007$	&	$60.5	\pm	1.2$	&		\\
$\nu_{24}$	&	$5.658204	\pm	0.000008$	&	$60.0	\pm	1.3$	&	 $\nu_{1} +  \nu_{3}$ 	 \\
$\nu_{25}$	&	$2.767057	\pm	0.000016$	&	$58.7	\pm	2.5$	&	 $-3\nu_{1}+   2\nu_{2}+   2\nu_{3}$ 	 \\
$\nu_{26}$	&	$5.505199	\pm	0.000008$	&	$53.8	\pm	1.2$	&	    $2\nu_{2}+   2\nu_{3}  -2\nu_{4}$ 	 \\
$\nu_{27	}$	&	$5.999424	\pm	0.000008$	&	$52.2	\pm	1.1$	&		\\
$\nu_{28	}$	&	$2.674685	\pm	0.000019$	&	$51.1	\pm	2.6$	&		\\
$\nu_{29}$	&	$5.893861	\pm	0.000011$	&	$41.9	\pm	1.3$	&	    $2\nu_{2}  - \nu_{3}+   \nu_{4}$ 	 \\
$\nu_{30	}$	&	$6.191502	\pm	0.000007$	&	$41.1	\pm	0.7$	&		\\
$\nu_{31	}$	&	$2.642377	\pm	0.000024$	&	$41.0	\pm	2.7$	&		\\
$\nu_{32}$	&	$2.714778	\pm	0.000024$	&	$39.5	\pm	2.5$	&	 $2\nu_{1}+   2\nu_{2}     -3\nu_{4}$ 	 \\
$\nu_{33	}$	&	$5.895022	\pm	0.000012$	&	$38.6	\pm	1.3$	&		\\
$\nu_{34	}$	&	$5.556988	\pm	0.000013$	&	$35.4	\pm	1.3$	&		\\
$\nu_{35	}$	&	$12.085640	\pm	0.000005$	&	$34.3	\pm	0.5$	&		\\
$\nu_{36	}$	&	$2.656498	\pm	0.000030$	&	$33.5	\pm	2.7$	&		\\
$\nu_{37}$	&	$5.850234	\pm	0.000015$	&	$33.2	\pm	1.3$	&	          $2\nu_{4}$ 	 \\
$\nu_{38	}$	&	$6.267179	\pm	0.000007$	&	$32.8	\pm	0.6$	&		\\
$\nu_{39	}$	&	$2.902992	\pm	0.000042$	&	$28.3	\pm	3.2$	&		\\
$\nu_{40	}$	&	$5.529417	\pm	0.000017$	&	$27.7	\pm	1.2$	&		\\
$\nu_{41}$	&	$2.733045	\pm	0.000033$	&	$27.2	\pm	2.4$	&	 $\nu_{1}+   \nu_{3} -\nu_{4}$ 	 \\
$\nu_{42	}$	&	$5.856770	\pm	0.000018$	&	$26.8	\pm	1.3$	&		\\
$\nu_{43	}$	&	$6.058380	\pm	0.000013$	&	$26.7	\pm	0.9$	&		\\
$\nu_{44}$	&	$5.663315	\pm	0.000018$	&	$26.0	\pm	1.3$	&	 $3\nu_{1}        - \nu_{4}$ 	 \\
$\nu_{45	}$	&	$5.762114	\pm	0.000019$	&	$25.7	\pm	1.3$	&		\\
$\nu_{46}$	&	$5.899607	\pm	0.000018$	&	$25.6	\pm	1.3$	&	 $2\nu_{1}+   2\nu_{2}  -2\nu_{3}$ 	 \\
$\nu_{47}$	&	$5.528646	\pm	0.000018$	&	$25.4	\pm	1.2$	&	 $\nu_{1}+      2\nu_{3}  -\nu_{4}$ 	 \\
$\nu_{48	}$	&	$5.651492	\pm	0.000019$	&	$25.3	\pm	1.3$	&		\\
$\nu_{49	}$	&	$6.003772	\pm	0.000016$	&	$25.1	\pm	1.1$	&		\\
$\nu_{50	}$	&	$5.685564	\pm	0.000020$	&	$24.1	\pm	1.3$	&		\\
$\nu_{51}$	&	$5.677776	\pm	0.000021$	&	$23.4	\pm	1.3$	&	    $\nu_{2} +  \nu_{3}$ 	 \\
$\nu_{52	}$	&	$5.769912	\pm	0.000021$	&	$22.7	\pm	1.3$	&		\\
$\nu_{53	}$	&	$5.797322	\pm	0.000022$	&	$22.5	\pm	1.3$	&		\\
$\nu_{54}$	&	$5.649452	\pm	0.000023$	&	$21.6	\pm	1.3$	&	 $-3\nu_{1}+   3\nu_{2}+   2\nu_{3}$ 	 \\
$\nu_{55}$	&	$2.598874	\pm	0.000048$	&	$21.1	\pm	2.7$	&	 $-\nu_{1}+      3\nu_{3}  -\nu_{4}$ 	 \\
$\nu_{56	}$	&	$5.433314	\pm	0.000023$	&	$20.2	\pm	1.2$	&		\\
$\nu_{57	}$	&	$2.616321	\pm	0.000051$	&	$20.2	\pm	2.8$	&		\\
$\nu_{58}$	&	$5.845043	\pm	0.000024$	&	$20.1	\pm	1.3$	&	 $-2\nu_{1}+   \nu_{3} +  3\nu_{4}$ 	 \\
$\nu_{59	}$	&	$5.690009	\pm	0.000024$	&	$19.4	\pm	1.3$	&		\\
$\nu_{60	}$	&	$2.574995	\pm	0.000053$	&	$18.9	\pm	2.7$	&		\\
\hline
\end{tabular}
\end{table}

\begin{table}
  \contcaption{} %Frequency list for KIC 5608334 }
%  \label{tab:freq}
  \begin{tabular}{cccccc}
    \hline
%    $\nu_i$ & Frequency & Frequency error & Amplitude & Amplitude error & parent modes \\
    $\nu_i$ & Frequency  & Amplitude  & parent modes \\
%     & [c/d] & [c/d] & [ppm] & [ppm] &  \\
      & [d$^{-1}$]  & [ppm]  &  \\
   \hline
$\nu_{61	}$	&	$5.587489	\pm	0.000025$	&	$18.7	\pm	1.2$	&		\\
$\nu_{62	}$	&	$5.801623	\pm	0.000027$	&	$18.5	\pm	1.3$	&		\\
$\nu_{63}$	&	$2.633054	\pm	0.000060$	&	$17.3	\pm	2.7$	&	 $4\nu_{1}+   \nu_{2}  -4\nu_{4}$ 	 \\
$\nu_{64}$	&	$5.503826	\pm	0.000026$	&	$17.0	\pm	1.2$	&	  $-\nu_{2}+   3\nu_{3}$ 	 \\
$\nu_{65	}$	&	$5.622639	\pm	0.000029$	&	$17.0	\pm	1.3$	&		\\
$\nu_{66	}$	&	$3.000129	\pm	0.000088$	&	$16.9	\pm	4.0$	&		\\
$\nu_{67	}$	&	$5.558977	\pm	0.000028$	&	$16.7	\pm	1.3$	&		\\
$\nu_{68	}$	&	$11.864608	\pm	0.000012$	&	$16.4	\pm	0.5$	&		\\
$\nu_{69}$	&	$8.473207	\pm	0.000015$	&	$16.3	\pm	0.6$	&	    $\nu_{2}+   2\nu_{3}$ 	 \\
$\nu_{70	}$	&	$5.561736	\pm	0.000029$	&	$16.1	\pm	1.3$	&		\\
$\nu_{71}$	&	$5.616327	\pm	0.000032$	&	$15.5	\pm	1.3$	&	 $\nu_{1}+   4\nu_{2}     -3\nu_{4}$ 	 \\
$\nu_{72}$	&	$5.846586	\pm	0.000031$	&	$15.4	\pm	1.3$	&	 $-2\nu_{1}+   3\nu_{2}+   \nu_{4}$ 	 \\
$\nu_{73}$	&	$5.745129	\pm	0.000037$	&	$14.0	\pm	1.4$	&	 $\nu_{1}+   \nu_{2}$ 	 \\
$\nu_{74}$	&	$5.974663	\pm	0.000028$	&	$13.9	\pm	1.1$	&	 $-2\nu_{1}+  4\nu_{4}$ \\
$\nu_{75	}$	&	$5.282056	\pm	0.000030$	&	$13.8	\pm	1.1$	&		\\
$\nu_{76}$	&	$5.803584	\pm	0.000038$	&	$13.1	\pm	1.3$	&	 $-2\nu_{1}+   4\nu_{2}$ 	 \\
$\nu_{77}$	&	$5.456832	\pm	0.000034$	&	$12.9	\pm	1.2$	&	 $-2\nu_{1}+    4\nu_{3}$ 	 \\
$\nu_{78}$	&	$5.643731	\pm	0.000039$	&	$12.7	\pm	1.3$	&	 $4\nu_{1}  -\nu_{2}     -\nu_{4}$ 	\\
$\nu_{79	}$	&	$5.409264	\pm	0.000036$	&	$12.2	\pm	1.2$	&		\\
$\nu_{80	}$	&	$5.315772	\pm	0.000036$	&	$12.0	\pm	1.2$	&		\\
$\nu_{81	}$	&	$8.834344	\pm	0.000019$	&	$11.7	\pm	0.6$	&		\\
$\nu_{82	}$	&	$5.349422	\pm	0.000039$	&	$11.4	\pm	1.2$	&		\\
$\nu_{83}$	&	$8.521015	\pm	0.000022$	&	$11.4	\pm	0.7$	&	 $2\nu_{1}+  \nu_{3}$ 	 \\
$\nu_{84}$	&	$5.562584	\pm	0.000043$	&	$11.1	\pm	1.3$	&	 $-3\nu_{1}+  2\nu_{2}+   3\nu_{3}$ 	 \\
$\nu_{85}$	&	$11.355527	\pm	0.000019$	&	$10.8	\pm	0.5$	&	    $2\nu_{2}+  2\nu_{3}$ 	 \\
$\nu_{86}$	&	$5.268977	\pm	0.000039$	&	$10.5	\pm	1.1$	&	 $\nu_{1}+   4\nu_{3}  -3\nu_{4}$ 	 \\
$\nu_{87	}$	&	$8.334188	\pm	0.000022$	&	$10.3	\pm	0.6$	&		\\
$\nu_{88}$	&	$8.909514	\pm	0.000021$	&	$10.1	\pm	0.6$	&	 $2\nu_{1}  -2\nu_{3}+   3\nu_{4}$ 	 \\
$\nu_{89	}$	&	$5.960207	\pm	0.000041$	&	$10.1	\pm	1.1$	&		\\
$\nu_{90}$	&	$5.461304	\pm	0.000044$	&	$10.0	\pm	1.2$	&	       $3\nu_{3}  -\nu_{4}$ 	 \\
$\nu_{91}$	&	$5.721355	\pm	0.000052$	&	$9.9	\pm	1.4$	&	    $3\nu_{2}  -\nu_{4}$ 	 \\
$\nu_{92	}$	&	$6.278255	\pm	0.000024$	&	$9.8	\pm	0.6$	&		\\
$\nu_{93	}$	&	$5.389924	\pm	0.000047$	&	$9.6	\pm	1.2$	&		\\
$\nu_{94	}$	&	$5.369354	\pm	0.000045$	&	$9.5	\pm	1.1$	&		\\
$\nu_{95	}$	&	$5.437901	\pm	0.000050$	&	$9.1	\pm	1.2$	&		\\
$\nu_{96	}$	&	$8.990661	\pm	0.000024$	&	$8.8	\pm	0.6$	&		\\
$\nu_{97	}$	&	$8.766174	\pm	0.000024$	&	$8.8	\pm	0.6$	&		\\
$\nu_{98	}$	&	$8.424134	\pm	0.000026$	&	$8.5	\pm	0.6$	&		\\
$\nu_{99	}$	&	$8.636562	\pm	0.000027$	&	$8.4	\pm	0.6$	&		\\
$\nu_{100}$	&	$11.286926	\pm	0.000025$	&	$8.3	\pm	0.6$	&		\\
$\nu_{101}$	&	$12.086311	\pm	0.000024$	&	$7.7	\pm	0.5$	&		\\
$\nu_{102}$	&	$8.144870	\pm	0.000030$	&	$7.7	\pm	0.6$	&		\\
$\nu_{103}$	&	$8.764023	\pm	0.000028$	&	$7.5	\pm	0.6$	&		\\
$\nu_{104}$	&	$8.906329	\pm	0.000030$	&	$7.2	\pm	0.6$	&	    $3\nu_{2}  -2\nu_{3}+   2\nu_{4}$ 	 \\
$\nu_{105}$	&	$8.762505	\pm	0.000029$	&	$7.2	\pm	0.6$	&	 $3\nu_{1}+   2\nu_{2} -2\nu_{3}$ 	 \\
$\nu_{106}$	&	$8.583393	\pm	0.000035$	&	$6.9	\pm	0.7$	&	 $\nu_{1}+      \nu_{3}+   \nu_{4}$ 	 \\
$\nu_{107}$	&	$11.768250	\pm	0.000031$	&	$6.6	\pm	0.6$	&	 $\nu_{1}     -\nu_{3}+   4\nu_{4}$ 	 \\
$\nu_{108}$	&	$11.971333	\pm	0.000032$	&	$6.4	\pm	0.5$	&		\\
$\nu_{109}$	&	$8.990056	\pm	0.000033$	&	$6.3	\pm	0.6$	&		\\
$\nu_{110}$	&	$8.333354	\pm	0.000035$	&	$6.3	\pm	0.6$	&		\\
$\nu_{111}$	&	$8.430330	\pm	0.000037$	&	$6.1	\pm	0.6$	&	    $2\nu_{2}+   2\nu_{3}  -\nu_{4 }$	 \\
$\nu_{112}$	&	$11.225740	\pm	0.000035$	&	$6.0	\pm	0.6$	&	    $2\nu_{2}+   3\nu_{3}  -\nu_{4 }$	 \\
$\nu_{113}$	&	$11.513384	\pm	0.000037$	&	$6.0	\pm	0.6$	&	 $3\nu_{1}+     \nu_{4}$ 	 \\
$\nu_{114}$	&	$11.427834	\pm	0.000035$	&	$5.8	\pm	0.5$	&	 $3\nu_{1}+   2\nu_{2}    -\nu_{4}$ 	 \\
$\nu_{115}$	&	$8.526019	\pm	0.000044$	&	$5.8	\pm	0.7$	&	 $4\nu_{1}     -\nu_{4}$ 	 \\
$\nu_{116}$	&	$11.671535	\pm	0.000037$	&	$5.7	\pm	0.6$	&		\\
$\nu_{117}$	&	$8.637887	\pm	0.000041$	&	$5.5	\pm	0.6$	&		\\
$\nu_{118}$	&	$8.700036	\pm	0.000039$	&	$5.4	\pm	0.6$	&		\\
$\nu_{119}$	&	$8.377842	\pm	0.000042$	&	$5.3	\pm	0.6$	&		\\
$\nu_{120}$	&	$8.986568	\pm	0.000042$	&	$5.2	\pm	0.6$	&		\\
\hline
\end{tabular}
\end{table}

\begin{table}
  \contcaption{} %Frequency list for KIC 5608334 }
%  \label{tab:freq}
  \begin{tabular}{cccccc}
    \hline
%    $\nu_i$ & Frequency & Frequency error & Amplitude & Amplitude error & parent modes \\
    $\nu_i$ & Frequency  & Amplitude  & parent modes \\
%     & [c/d] & [c/d] & [ppm] & [ppm] &  \\
      & [d$^{-1}$]  & [ppm]  &  \\
   \hline
$\nu_{121}$	&	$11.429624	\pm	0.000040$	&	$5.2	\pm	0.5$	&		\\
$\nu_{122}$	&	$11.361886	\pm	0.000039$	&	$5.1	\pm	0.5$	&		\\
$\nu_{123}$	&	$9.077137	\pm	0.000043$	&	$5.0	\pm	0.6$	&		\\
$\nu_{124}$	&	$11.052332	\pm	0.000042$	&	$5.0	\pm	0.6$	&		\\
$\nu_{125}$	&	$11.433447	\pm	0.000041$	&	$5.0	\pm	0.5$	&		\\
$\nu_{126}$	&	$11.766748	\pm	0.000041$	&	$4.9	\pm	0.5$	&		\\
$\nu_{127}$	&	$8.481080	\pm	0.000050$	&	$4.9	\pm	0.7$	&		\\
$\nu_{128}$	&	$8.837499	\pm	0.000045$	&	$4.8	\pm	0.6$	&	 $-\nu_{1}+    4\nu_{4}$ 	 \\
$\nu_{129}$	&	$11.587226	\pm	0.000043$	&	$4.8	\pm	0.6$	&		\\
$\nu_{130}$	&	$11.167284	\pm	0.000043$	&	$4.7	\pm	0.5$	&		\\
$\nu_{131}$	&	$8.474172	\pm	0.000051$	&	$4.7	\pm	0.6$	&	    $4\nu_{2}+   \nu_{3}  -2\nu_{4}$ 	 \\
$\nu_{132}$	&	$9.083654	\pm	0.000046$	&	$4.6	\pm	0.6$	&		\\
$\nu_{133}$	&	$11.289021	\pm	0.000046$	&	$4.5	\pm	0.6$	&		\\
$\nu_{134}$	&	$8.481808	\pm	0.000055$	&	$4.5	\pm	0.7$	&	 $4\nu_{1} -2\nu_{2}+   \nu_{3}$ 	 \\
$\nu_{135}$	&	$8.251661	\pm	0.000050$	&	$4.5	\pm	0.6$	&		\\
$\nu_{136}$	&	$8.646842	\pm	0.000052$	&	$4.3	\pm	0.6$	&	    $3\nu_{2}$ 	 \\
$\nu_{137}$	&	$11.227152	\pm	0.000050$	&	$4.3	\pm	0.6$	&		\\
$\nu_{138}$	&	$8.632684	\pm	0.000054$	&	$4.2	\pm	0.6$	&		\\
$\nu_{139}$	&	$11.769328	\pm	0.000049$	&	$4.2	\pm	0.6$	&		\\
$\nu_{140}$	&	$8.425461	\pm	0.000056$	&	$4.0	\pm	0.6$	&	 $-2\nu_{1}+   2\nu_{2}+   3\nu_{3}$ 	 \\
$\nu_{141}$	&	$8.379683	\pm	0.000055$	&	$3.9	\pm	0.6$	&		\\
$\nu_{142}$	&	$8.381844	\pm	0.000056$	&	$3.9	\pm	0.6$	&	 $-2\nu_{1}+  4\nu_{3}+  \nu_{4}$ 	 \\
$\nu_{143}$	&	$8.054455	\pm	0.000066$	&	$3.7	\pm	0.7$	&		\\
$\nu_{144}$	&	$8.291565	\pm	0.000060$	&	$3.7	\pm	0.6$	&		\\
$\nu_{145}$	&	$11.163591	\pm	0.000055$	&	$3.7	\pm	0.5$	&		\\
$\nu_{146}$	&	$11.001182	\pm	0.000057$	&	$3.7	\pm	0.6$	&		\\
$\nu_{147}$	&	$8.292629	\pm	0.000061$	&	$3.7	\pm	0.6$	&		\\
$\nu_{148}$	&	$8.833616	\pm	0.000059$	&	$3.7	\pm	0.6$	&	 $-3\nu_{1}+   3\nu_{2}+      3\nu_{4}$ 	 \\
$\nu_{149}$	&	$11.772323	\pm	0.000057$	&	$3.6	\pm	0.6$	&	 $3\nu_{1}     -2\nu_{3}+   3\nu_{4}$ 	 \\
$\nu_{150}$	&	$11.216446	\pm	0.000057$	&	$3.6	\pm	0.5$	&		\\
$\nu_{151}$	&	$11.759458	\pm	0.000057$	&	$3.6	\pm	0.5$	&		\\
$\nu_{152}$	&	$12.205213	\pm	0.000052$	&	$3.5	\pm	0.5$	&		\\
$\nu_{153}$	&	$8.588264	\pm	0.000068$	&	$3.5	\pm	0.6$	&	 $3\nu_{1}$ 	 \\
$\nu_{154}$	&	$11.107640	\pm	0.000057$	&	$3.5	\pm	0.5$	&		\\
$\nu_{155}$	&	$8.838954	\pm	0.000062$	&	$3.5	\pm	0.6$	&		\\
$\nu_{156}$	&	$11.509859	\pm	0.000064$	&	$3.4	\pm	0.6$	&	 $\nu_{1}+   3\nu_{2}$ 	 \\
$\nu_{157}$	&	$8.842500	\pm	0.000064$	&	$3.4	\pm	0.6$	&	 $\nu_{1}  -\nu_{3}+   3\nu_{4}$ 	 \\
$\nu_{158}$	&	$8.698089	\pm	0.000062$	&	$3.4	\pm	0.6$	&		\\
$\nu_{159}$	&	$11.872966	\pm	0.000061$	&	$3.3	\pm	0.5$	&		\\
$\nu_{160}$	&	$8.342010	\pm	0.000067$	&	$3.3	\pm	0.6$	&	   $-2\nu_{2}+   4\nu_{3}+   \nu_{4}$ 	 \\
$\nu_{161}$	&	$11.506292	\pm	0.000065$	&	$3.3	\pm	0.6$	&		\\
$\nu_{162}$	&	$11.050652	\pm	0.000065$	&	$3.2	\pm	0.6$	&		\\
$\nu_{163}$	&	$8.213138	\pm	0.000071$	&	$3.2	\pm	0.6$	&		\\
$\nu_{164}$	&	$11.500441	\pm	0.000067$	&	$3.2	\pm	0.6$	&		\\
$\nu_{165}$	&	$8.255114	\pm	0.000069$	&	$3.2	\pm	0.6$	&		\\
$\nu_{166}$	&	$8.959990	\pm	0.000070$	&	$3.2	\pm	0.6$	&		\\
$\nu_{167}$	&	$11.106039	\pm	0.000064$	&	$3.1	\pm	0.5$	&		\\
$\nu_{168}$	&	$11.676005	\pm	0.000068$	&	$3.1	\pm	0.6$	&		\\
$\nu_{169}$	&	$11.056776	\pm	0.000069$	&	$3.0	\pm	0.5$	&	 $2\nu_{1}+ 4\nu_{3} -2\nu_{4}$ 	 \\
$\nu_{170}$	&	$8.118941	\pm	0.000082$	&	$3.0	\pm	0.7$	&		\\
$\nu_{171}$	&	$8.689643	\pm	0.000072$	&	$3.0	\pm	0.6$	&	  $2\nu_{2}+ \nu_{4}$ 	 \\
$\nu_{172}$	&	$11.158671	\pm	0.000069$	&	$2.9	\pm	0.5$	&		\\
$\nu_{173}$	&	$8.296201	\pm	0.000076$	&	$2.9	\pm	0.6$	&		\\
$\nu_{174}$	&	$8.856920	\pm	0.000075$	&	$2.9	\pm	0.6$	&	 $-2\nu_{1}+  \nu_{2}+ 4\nu_{4}$ 	 \\
$\nu_{175}$	&	$8.650534	\pm	0.000075$	&	$2.9	\pm	0.6$	&	 $2\nu_{1}+    \nu_{4}$ 	 \\
$\nu_{176}$	&	$11.588657	\pm	0.000071$	&	$2.9	\pm	0.6$	&		\\
$\nu_{177}$	&	$8.113250	\pm	0.000085$	&	$2.9	\pm	0.6$	&		\\
$\nu_{178}$	&	$8.148755	\pm	0.000083$	&	$2.8	\pm	0.6$	&		\\
$\nu_{179}$	&	$11.697220	\pm	0.000076$	&	$2.8	\pm	0.6$	&	 $-2\nu_{1}+  3\nu_{2}+ 3\nu_{4}$ 	 \\
$\nu_{180}$	&	$9.073512	\pm	0.000076$	&	$2.8	\pm	0.6$	&		\\
\hline
\end{tabular}
\end{table}

\begin{table}
  \contcaption{} %Frequency list for KIC 5608334 }
%  \label{tab:freq}
  \begin{tabular}{cccccc}
    \hline
%    $\nu_i$ & Frequency & Frequency error & Amplitude & Amplitude error & parent modes \\
    $\nu_i$ & Frequency  & Amplitude  & parent modes \\
%     & [c/d] & [c/d] & [ppm] & [ppm] &  \\
      & [d$^{-1}$]  & [ppm]  &  \\
   \hline
$\nu_{181}$	&	$10.856372	\pm	0.000074$	&	$2.8	\pm	0.5$	&		\\
$\nu_{182}$	&	$10.905030	\pm	0.000074$	&	$2.8	\pm	0.6$	&		\\
$\nu_{183}$	&	$11.502347	\pm	0.000077$	&	$2.8	\pm	0.6$	&		\\
$\nu_{184}$	&	$11.046784	\pm	0.000076$	&	$2.7	\pm	0.5$	&		\\
$\nu_{185}$	&	$11.763508	\pm	0.000076$	&	$2.7	\pm	0.5$	&		\\
$\nu_{186}$	&	$10.903098	\pm	0.000077$	&	$2.7	\pm	0.6$	&		\\
$\nu_{187}$	&	$11.863159	\pm	0.000078$	&	$2.6	\pm	0.5$	&		\\
$\nu_{188}$	&	$11.511758	\pm	0.000083$	&	$2.6	\pm	0.6$	&		\\
$\nu_{189}$	&	$11.108837	\pm	0.000077$	&	$2.6	\pm	0.5$	&		\\
$\nu_{190}$	&	$11.498292	\pm	0.000085$	&	$2.5	\pm	0.6$	&		\\
$\nu_{191}$	&	$10.770712	\pm	0.000085$	&	$2.5	\pm	0.6$	&		\\
$\nu_{192}$	&	$11.306584	\pm	0.000083$	&	$2.5	\pm	0.5$	&	 $-2\nu_{1}+ 4\nu_{3}+ 2\nu_{4}$ 	 \\
\hline
\end{tabular}
%\end{longtable}
 \end{table}

%%%%%%%%%%%%%%%%%%%%%%%%%%%%%%%%%%%%%%%%%%%%%%%%%%

% Don't change these lines
\bsp	% typesetting comment
\label{lastpage}
\end{document}